# Non-Reciprocity in Photon Polarization based on direction of polarizer under Gravitational Fields

Hansol Noh[1,2], Paul M. Alsing[3], Warner A. Miller[3], and Doyeol Ahn[1,3]*,

[1]*Department of Electrical and Computer Engineering and Center for Quantum Information Processing, University of Seoul, 163 Seoulsiripdae-ro, Dongdaemun-gu, Seoul, 02504, Republic of Korea*

[2]*Department of Electrical and Computer Engineering, Seoul National University Seoul 08826, Republic of Korea.*

[3]*Department of Physics, Florida Atlantic University, 777 Glades Road, Boca Rat Raton, FL 33431-0991, USA.*

*\*Correspondence to: <u>dahn@uos.ac.kr</u>.*

Abstract

Unification of gravity with quantum mechanics is still a terra incognita. Photon polarization measurements offer a unique window for probing the interaction between these two fundamental forces. We have revealed that non-reciprocity in the photon polarization angle can arise by tailoring the quantization axis, which corresponds to the direction of polarizer. Due to this non-reciprocity, the measured polarization angle can become ten times larger than that of gravitationally induced frame rotation in both near-Earth and black hole environments. To verify this finding, we propose an astronomical interferometer composed of satellites with the tailored quantization axis, challenging the conventional view of their triviality in closed paths of a photon. Notably, this non-reciprocity can extend to any rotation in the polarization plane, irrespective of the origins, all of which can dictate polarization rotation. Our findings could offer new opportunities for testing fundamental principles in physics.

# I. INTRODUCTION

The unification of general relativity and quantum mechanics remains one of the most significant challenge in theoretical physics. General relativity governs the large-scale dynamics of the universe and quantum mechanics dictates the behavior of quantum particles at Planck scales. However, the description of quantum particles under gravitational fields has not been fully understood. Seminal works such as Hawking radiation[1] and Unruh effect[2], have provided some insights into the interaction between quantum particles and classical gravitational field. Nevertheless, complete understanding of the interplay between two pillars of physics is still lacking, leaving ample room for further discoveries[3].

Photons, as inherently massless entities, move along null geodesics, positioning them as sensitive probes of spacetime curvature. The essential roles of photons in advanced experimental systems from interferometers[4-6] to quantum communication systems[7-23], further emphasize the significance of comprehending the nature of photon states under gravitational influence. Moreover, state-of-the-art advancements in high-precision measurements of relativistic effects, such as the Laser Interferometer Space Antenna (LISA) [5], and, like Gravity Probe B[4], have made probing gravity's subtle influence on light increasingly feasible. A recent study employing the Micius satellite examined the decoherence effects of entangled photons in Earth's gravitational field, providing data that challenges the predictions of event formalism and is consistent with the standard quantum formalism[24]. These technological developments open new arenas for investigating the intricate interplay between general relativity and quantum mechanics, furthering our understanding of the universe's fundamental nature.

Furthermore, the polarization of a photon provides a unique window for probing various advanced fields such as quantum gravity[25,26], potential violations of Einstein equivalence principle[27,28], and broader areas of astrophysics[29] and cosmology[30], as the polarization changes when the photon moves through spacetime. Nevertheless, while the effects of time dilation have been widely explored, the intricacies of the spin (or polarization) states of a photon in spacetime have been underexplored.

In flat spacetime, the exploration of photon polarization states has yielded key insights, notably into the Wigner rotation — an additional relativistic rotation arising from the non-commutative nature of Lorentz transformations[31,32]— and its associated quantum phase, the Wigner rotation angle (WRA), which becomes observable with spin states of quantum particles and elucidates how a spin state of a quantum particle interacts with the additional rotation. In this context, introducing a standard frame — where the wave vector is aligned to the third axis (called the quantization axis corresponding to the orientation of a polarizer) — is essential for understanding how quantum particles respond to Lorentz transformations between inertial frames[32]. It is employed to construct irreducible unitary representations of a single Lorentz transformation. In this frame, the representation decomposes into three sequential Lorentz transformations: from the standard frame to the original, an arbitrary Lorentz transformation in the original, and back to the standard frame. From non-commutativity of these three sequential Lorentz transformations, the Wigner rotation angle (WRA) arises naturally.

The effect of Wigner rotation in flat spacetime on quantum information has been widely studied. Previous study has shown that WRA affects entanglement between the spins of a pair of particles through spin-momentum mixing in moving frames [33]. Furthermore, it has been shown that momentum-dependent phases (WRA) result in variations in entanglement between a pair of polarization-entangled photon beams in moving frames depending on the boost direction[34]. Wigner rotation is introduced in both studies to study entanglement of quantum particles observed in moving frames and demonstrate that the WRA can lead to frame-dependent quantum effects.

The phenomenon becomes more intricate in curved spacetime. Employing Einstein equivalence principle and tetrad fields — a set of four orthonormal basis vectors that delineate spacetime curvature into locally flat segments, providing a means to describe various local inertial frames (corresponding to observers in different states of motion: e.g. stationary, freely falling, rotating, etc.) — the WRA can be extended into the geometric phase of photon helicity states in curved spacetime. In this framework, spacetime curvature is encoded in the variations in local inertial frames which manifest as local Lorentz transformations[35]. Naturally, their irreducible unitary representations and the associated Wigner rotation angles are employed in the quantum field description of local Lorentz transformations as a particle moves in curved spacetime.

For an open path of a photon, in Schwarzschild spacetime, there is a consensus that the phase is zero on open paths only with the proper choice of gauge[36]. Prior research has accentuated the gravitationally induced WRA can have values of $O(10^{-5})$ degrees, especially for observers in circular orbits of Earth (e.g. satellites). While these findings underscore the observational potential of WRA, especially in Earth's vicinity[37,38], the WRA of a photon traveling along a closed loop has conventionally been considered trivial, interpreted as a Berry phase influenced by the holonomy group of classical polarization vectors[39]. In Schwarzschild spacetime, the WRA is known to be zero on closed paths regardless of gauge[36]. For Kerr spacetime, it has been reported that a gauge-invariant result can be obtained using two future-directed trajectories whose layout is similar to the two arms of a Mach-Zehnder interferometer[36]. However, given that the effect of the Earth's spinning angular momentum is negligible, measurement via interferometers has been regarded as infeasible, despite the small values of WRA.

In this study, we explore the effects of the choice of quantization axis, which corresponds to the orientation of a polarizer and has not been previously considered, on the geometric phase (WRA) of a photon helicity state, leading to momentum-dependence in the WRA, for both flat and curved spacetime scenarios. Since the WRA depends on the Lorentz transformations rather than the frame, the choice of various possible standard frames can yield different values for the WRA, resulting in momentum-dependence in WRA. We find that the interplay between the tailored choices of the quantization axis and spacetime curvature results in non-reciprocity of the WRAs. This leads to measurable phase differences in astronomical interferometry conducted near Earth. Also, it is found that the conventional view of trivial WRA on a closed path is recovered with the quantization axis chosen to be parallel to the momentum-vector of a photon.

We propose employing two types of astronomical interferometers with a specifically tailored quantization axis to investigate the interactions between quantum spin states and classical gravitational fields. The first type, the combination of Hong–Ou–Mandel (HOM) and Mach-Zehnder interferometer is appreciated for its precision, particularly its sensitivity and its ability to resist noise from single-photon detection. Through monitoring the coincidence counts, the subtle geometric phase variances induced by spacetime curvature should be measurable. In parallel, we consider a Mach-Zehnder interferometer using a nearly monochromatic light source or a single-photon source. This setup is introduced for validating the equivalence between the Wigner rotation of photon states and the transformation of classical electric fields by comparing the results of the case with the single-photon source, which highlights photon state non-locality.

It is noteworthy that the non-reciprocity arising from the choice of quantization axis can extend beyond the WRA induced by local inertial frame variations. Establishing a standard frame aligned with the chosen quantization axis introduces a sequence of three Lorentz transformations including any arbitrary rotation regardless of its origin: starting from the standard frame, applying the rotation, and then returning to the original frame. Also, It is numerically shown that WRA can be ten times larger than the frame's rotation angle in both near-Earth and black hole environments by harnessing the non-reciprocity. This suggests that the observed non-reciprocity could be utilized in measuring subtle polarization rotations under varying gravitational conditions, opening doors for further exploration in how light and electromagnetic phenomena behave in different cosmic settings.

## II. RESULTS

### A. Theoretical Background

In the study of photon helicity states in spacetime, the standard frame is indispensable, wherein the third axis is designated as the quantization axis. Within this framework, the photon undergoes three sequential Lorentz transformations[32]; starting from the Lorentz transformation aligning the photon's wave vector in the standard frame to the transmission direction of a photon, a local Lorentz transformation induced by spacetime curvature, and finally returning to alignment with the quantization axis for comparison with the initial state. The non-commutative nature of the three Lorentz transformations gives rise to the additional rotation, called Wigner Rotation Angle (WRA), underscoring the significant role the quantization axis selection plays in determining the WRA.

Transformation between two local inertial frames, through which a photon passes, is described by an infinitesimal variation of the local orthonormal bases, called tetrads $e_{\hat{a}}{}^{\mu}(x)$, along an infinitesimal displacement of the photon, which, spanning the local inertial frames, satisfy the orthonormal constraints,

$$g_{\mu\nu}(x) = \eta_{\hat{a}\hat{b}} e^{\hat{a}}{}_{\mu}(x) e^{\hat{b}}{}_{\nu}(x); \; \hat{a} \text{ and } \mu = 0,1,2,3 \qquad (1)$$

with $\eta_{\hat{a}\hat{b}} = (-1,1,1,1)$. Considering Einstein equivalence principle, the variation of local inertial frames can be viewed as a local Lorentz transformation[35,40-43]. When a photon state moves along a null geodesic in the geometric optics limit[44], photon four-momentum components measured in local frames are transformed under an infinitesimal change of tetrads such that

$$k_{\hat{a}}(x) \rightarrow k'_{\hat{a}}(x) \equiv k_{\hat{a}}(x) + \delta k_{\hat{a}}(x) = \left( \delta_{\hat{a}}{}^{\hat{b}} + \lambda_{\hat{a}}{}^{\hat{b}}(x) d\xi \right) k_{\hat{b}}(x) = \Lambda_{\hat{a}}{}^{\hat{b}}(x) k_{\hat{b}}(x) \;, \qquad (2)$$

where $\lambda_{\hat{a}}{}^{\hat{b}}(x) = \left( \nabla_{\mathbf{k}} e_{\hat{a}}{}^{\nu}(x) \right) e_{\nu}{}^{\hat{b}}(x)$. The infinitesimal variation rate $\lambda_{\hat{a}}{}^{\hat{b}}(x)$ is antisymmetric[35,40] and thus can be naturally interpreted as an infinitesimal local Lorentz transformation, $\Lambda_{\hat{a}}{}^{\hat{b}}(x) = \delta_{\hat{a}}{}^{\hat{b}} + \lambda_{\hat{a}}{}^{\hat{b}}(x)$ [35,40,42,43] with Kronecker delta $\delta_a^{\ \hat{b}}$. A unitary representation of arbitrary Lorentz transformation, $\Lambda$, for a photon state with the helicity, $\sigma$, is described by[32]

$$U(\Lambda)\left|\hat{\mathbf{k}},\sigma\right\rangle = \sqrt{\frac{(\Lambda k)^{\hat{0}}}{k^{\hat{0}}}}\sum_{\sigma'}D_{\sigma'\sigma}\left(W(\Lambda,\hat{k})\right)\left|\hat{\mathbf{k}}_{\Lambda},\sigma'\right\rangle, \tag{3}$$

where the $W(\Lambda,\hat{k})$ and $D(W)$ represent the Wigner's little group element and its irreducible representation, respectively. Hereafter, the bolded letters indicate the spatial components of the four vector: $\hat{\mathbf{k}}$ refers to the spatial components of the wave vector in the local frame and $\hat{\mathbf{k}}_\Lambda$ signifies the Lorentz-transformed spatial components of the wave vector in the local frame. The necessity of Wigner's little group in the quantization process lies in its role in steering transformations of particles' internal degrees of freedom under Lorentz transformations. To formulate a detailed description of the system, it is essential to account for transformations possibly altering the particles' spin states while preserving momentum invariance. The Wigner's little group is defined as the subgroup of Poincaré group, which leaves the particle's four-momentum invariant in a specific reference frame — either the rest frame for a massive particle or the standard frame for a massless one. Accordingly, for photons, the little group $W$ is defined to leave the wave vector invariant in the standard frame conventionally defined as having the photon wave vector aligned along the third axis such that the photon 4-momentum becomes $\hat{k}_{\text{std}} = (1,0,0,1)$. As such, the WRA in Eq. (3) is defined as $W = L^{-1}(\Lambda\hat{k})\Lambda L(\hat{k})$. Here, $L(\hat{k})$ represents Lorenz transformation from $\hat{k}_{\text{std}}$ to $\hat{k}$. The non-commutativity of these sequential Lorentz transformations naturally leads to the WRA.

By translating the infinitesimal variations of tetrads into local Lorentz transformations between local inertial frames, the phase of a photon state evolves by an infinitesimal Wigner rotation angle $d\psi$ (IWRA) whose ratio with respect to an affine parameter $\xi$ is given by[37]

$$\frac{d\psi}{d\xi} = \lambda^{\hat{1}}_{\ \hat{2}} + \left[\frac{n^{\hat{1}}}{1+n^{\hat{3}}}\left(-\lambda^{\hat{0}}_{\ \hat{2}} + \lambda^{\hat{2}}_{\ \hat{3}}\right) + \frac{n^{\hat{2}}}{1+n^{\hat{3}}}\left(\lambda^{\hat{0}}_{\ \hat{1}} + \lambda^{\hat{3}}_{\ \hat{1}}\right)\right] = \tilde{\psi}_{\text{inf}}^{\text{geodetic}} + \psi_{\text{inf}}^{\text{residual}}, \tag{4}$$

where $n^i$ is defined as $k^i/|k^i|$ with $\hat{i}=1$, 2, and 3. The first term, $\lambda^{\hat{1}}_{\hat{2}}$, in Eq. (4) is the rotation in the $\hat{1}-\hat{2}$ plane about the local third axis, $\hat{3}$-axis, and the residual terms in the square bracket represents a momentum-dependent phase which we designate as $\tilde{\psi}^{\text{residual}}_{\text{inf.}}$. The first (classical) term corresponds to the rotation of the polarization about the $\hat{3}$-axis which should be disentangled from other rotations when considering WRA induced by the non-commutativity of local Lorentz transformation. Also, by setting the local third axis to be orthogonal to the observers' planes, the contribution from the geodetic precession can be isolated in the first term $\lambda^{\hat{1}}_{\hat{2}} = \tilde{\psi}^{\text{geodetic}}_{\text{inf.}}$. The total Wigner rotation angle $\psi$ can be obtained via a time-ordered integration of the IWRAs over the geodesic trajectory $x(\xi)$ of the photon such that

$$e^{i\psi(\Lambda,\vec{n})} = e^{i\psi^{\text{geodetic}}(\Lambda,\vec{n})+i\psi^{\text{residual}}(\Lambda,\vec{n})}$$
$$= T\exp\left[i\int\tilde{\psi}^{\text{geodetic}}_{\text{inf.}}(\Lambda(x(\xi)),n^i(\xi))d\xi\right] + T\exp\left[i\int\tilde{\psi}^{\text{residual}}_{\text{inf.}}(\Lambda(x(\xi)),n^i(\xi))d\xi\right] \quad (5)$$

where $T$ represents the time-ordering operator.

B. WRA for two open paths of a photon: Earth-satellite communication and a spinning black hole

We explore the dependency of the Wigner Rotation Angle (WRA) on the quantization axis by examining two distinct scenarios depicted in Figure 1: ground-station-to-satellite communication (Fig.1A), and orbiting observers in the equatorial plane of a rotating black hole (Fig.1B). Fig. 1A illustrates a satellite in the equatorial (yellow line) and polar (light blue line) orbits around Earth. In Fig.1B, yellow circles represent observers in an equatorial orbit around the black hole such as materials in an accretion disk, stars, or spacecrafts. The black hole scenario is chosen to study WRA in a context where such effects from the dependence are more pronounced. The three sequential Lorentz transformations in these scenarios can be interpreted as follows: prior to photon emission, the polarization or phase of the helicity state is measured in the standard frame where the wave vector is aligned with the local third axis $e_\phi$, called the quantization axis (Fig.1C) and rotated back to the original wave vector (Fig.1D). Then, as the photons propagate, photons experience local Lorentz transformations induced by the gravitational field. Upon reaching the receiver, the polarization and phase are measured again in the standard frame (Fig.1E). For brevity, we focus on the case where photon lies in the equatorial plane only. Throughout this paper, a photon field on a curved spacetime is assumed to have a spinor structure[32,45] and the $(-+++)$ metric signature is used, and the hatted letter represents local flat spacetime. Details of tetrads and photon's trajectory in a Kerr spacetime can be found in the Method section.

To isolate contribution from the geodetic precession, another relativistic effect, we intentionally choose the quantization axis to be orthogonal to the orbit plane[37]. We then examine how the Wigner Rotation Angle (WRA) changes with different orientations of the quantization axis relative to the wave vector. For this, we consider various ratios of $rk^\phi$ to $k^r$ — approximately the impact factors $b_{ph}$ of photon's trajectories[42,46]— for both cases; for the earth-satellite case, the photon is sent off with the different ratios $rk^\phi/k$ at the radius of Earth, and for the black hole (BH), photons are sent from an observer at a distance of 4.5 times the Schwarzschild radius from the BH with the various ratios $rk^\phi/k$.

In our analysis of satellites in polar orbits, we noted that the signs of the ratios of $rk^\phi$ to $k^r$ significantly influence the WRA, as illustrated in Fig.1F; WRAs for the negative ratios have ten times larger values than those for the positive. Also, for equatorial orbits, while WRA variations are not discernible for the Earth case, in black hole orbit scenarios, the impact of sign change is apparent for WRAs, indicating a complex interplay with the sign of the impact parameter $b_{ph}$ on both orbits in general. The WRAs calculated for observers in equatorial plane of M87* black hole are presented in Fig. 1G.

The observed discrepancy between the two different orbits near Earth is attributed to the momentum-dependence of WRA, i.e., the dependence on the relative direction of the quantization axis compared to the wave vector. Central to this dependence is the choice of the quantization axis, which determines the two Lorentz transformations — the initial transformation from the standard frame to the original frame and the subsequent return transformation to the standard frame after the local Lorentz transformation induced by spacetime curvature. Such intricacies hold significant implications for the polar orbit case, where both wave vectors and quantization axis lie in the same plane. In such a case, when a local Lorentz transformation is applied, WRA changes depending on the sign of $\hat{z}$ - component of the wave vector, $n^{\hat{3}}$, while the direction of the $\hat{z}$-component wave vector $n^{\hat{3}}$ alone without any Lorentz transformation does not affect the Wigner rotation angle.

For the specific case illustrated in Fig.2, as an example, we consider the case where photon momentum lies in $\hat{y}\hat{z}$-plane of a local frame. In the absence of any rotation, WRA is null as illustrated in Figs.2A-E. Also, we consider the situation where the local frame undergoes the rotation $R_{\hat{y}}(\Delta\phi)$ about the $\hat{y}$-axis by the angle $\Delta\phi$, the corresponding rotation $R^{-1}(\hat{\mathbf{k}}')$ — aligning the rotated wave vector $\hat{\mathbf{k}}' = R_{\hat{y}}(\Delta\phi)\hat{\mathbf{k}}$ to the standard wave vector $\hat{k}_{\text{std}} = (1,0,0,1)$ — is directly influenced by the relative orientation of the quantization axis ($\hat{z}$-axis, marked by purple arrows in Fig.2) with respect to the wave vector $\hat{\mathbf{k}}'$, as per the definition of the rotation $R^{-1}(\hat{\mathbf{k}}')$. This interrelation leads to asymmetry in WRA depending on the relative orientation of the quantization axis as well as $R_{\hat{y}}(\Delta\phi)$, as depicted in Figs. 2-J. For infinitesimal $\hat{y}$-axis rotation angle $\Delta\phi$, the corresponding WRA is $\text{Arg}\left[1 - \Delta\phi n^{\hat{2}} / \left(0.5n^{\hat{2}}\Delta\phi + \mathrm{i}\left(1 + n^{\hat{3}}\right)\right)\right]$.

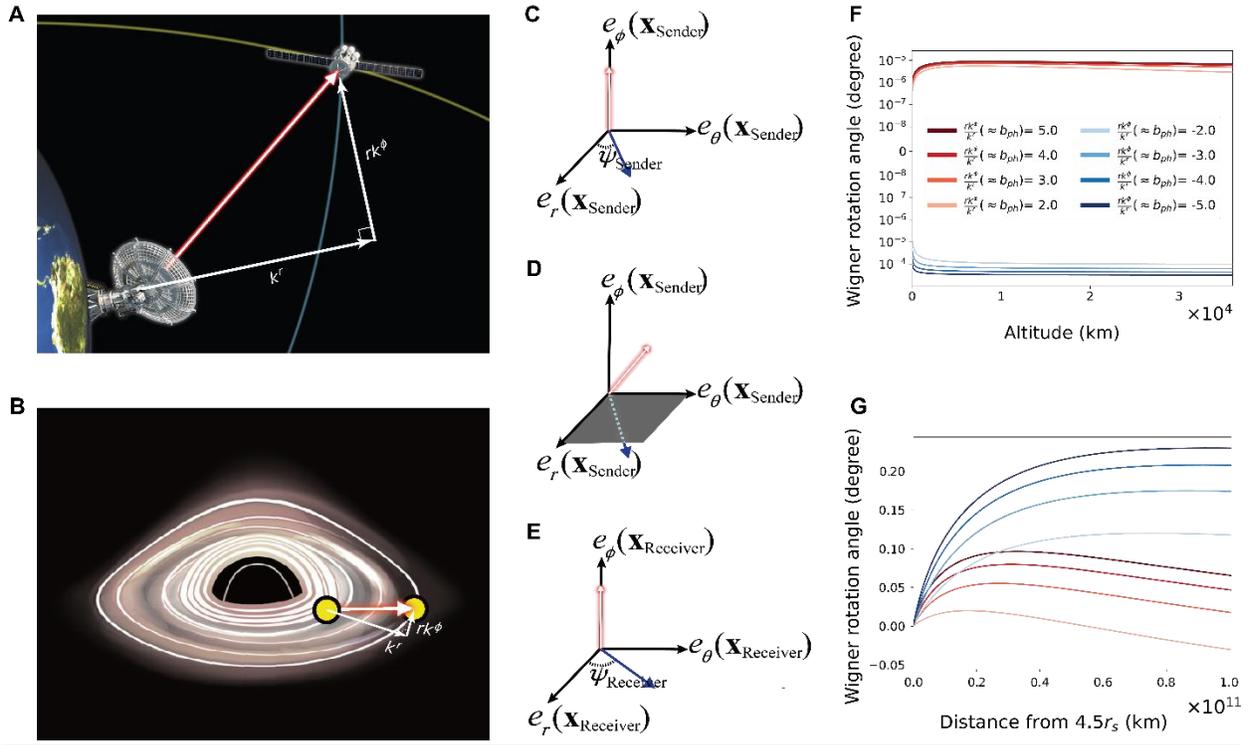

**Figure 1** Ground-station-to-satellite (Fig.1A), and orbiting observers in an equatorial plane of a rotating black hole (Fig.1B). Yellow and light blue lines in Fig.1A represent an equatorial and polar orbit, respectively. Yellow circles in Fig.1B represent observers orbiting around the black hole; we consider the case where photons are emitted from 4.5 times the Schwarzschild radius ($r_s$) away from the black hole. For both cases, various ratios $rk^\phi$ to $k^r$ are considered. As depicted in Figs.1C-E, wave vectors of photons are aligned with the local third axis, $e_\phi$, i.e., transformed to the standard frame. Before sending a photon, the polarization (i.e. the phase of helicity state) is measured in the standard frame (Fig.1C). Then, photons are rotated due to the WRA induced by gravity as depicted in Fig.1D. At the receiver, polarization and phase are again measured in the standard frame. While both WRAs for the equatorial and polar-orbit cases have the dependence on the impact factors of photon trajectories, the case of observers in the equatorial orbits near the Earth does not give discernible difference due to the weak gravity. For the cases of Earth-satellite-in-polar-orbits and observers in the equatorial plane of M87* black hole, WRAs are plotted as in Figs.1F and 1G, respectively. The original image of Earth used in Fig.1A is obtained from Yeongkwang Kim[47].

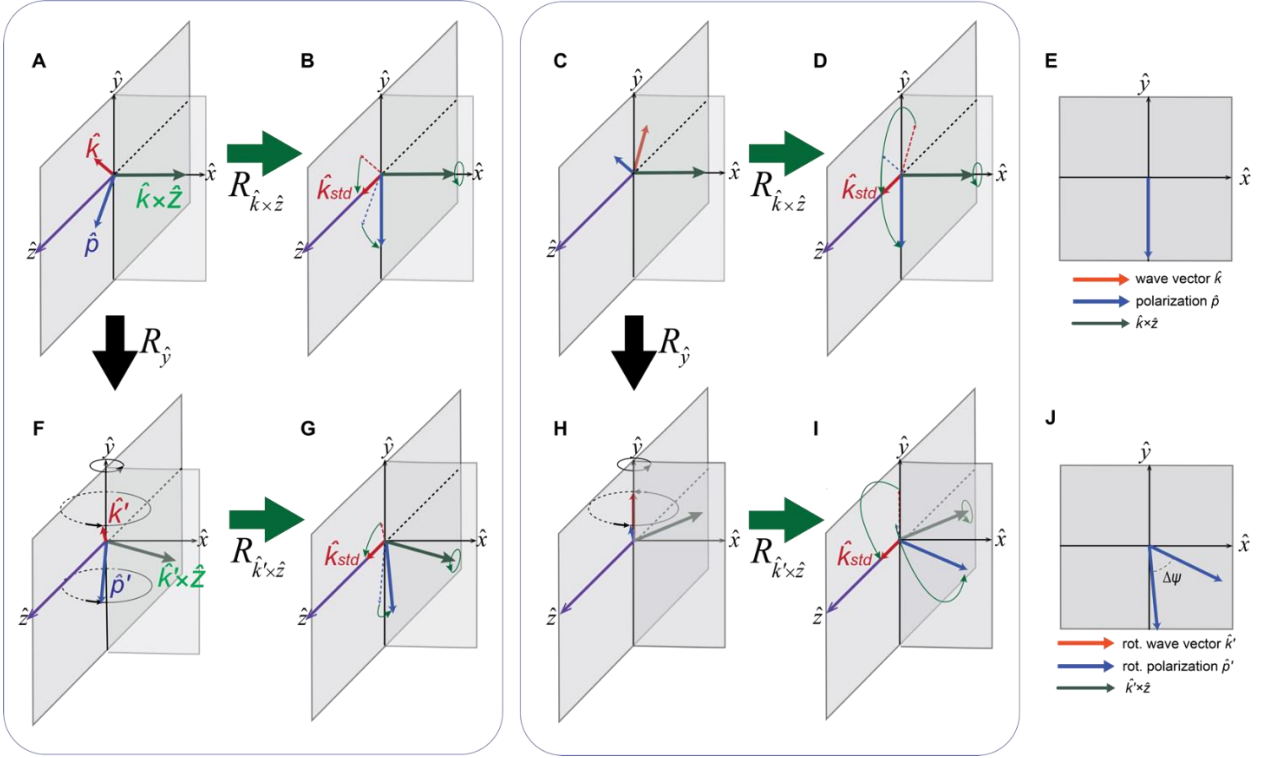

**Figure 2** Dependence of WRA on the choice of quantization axis. Using the $\hat{z}$-axis as the quantization axis and the photons on the $\hat{y}\hat{z}$-plane as an example. Wave vector $\hat{k}$ and polarization vector $\hat{p}$ are illustrated by red and blue arrows, respectively (Fig.2A). In the absence of any Lorentz-induced rotation, when the wave vector's $\hat{z}$-component aligns with the quantization axis, the $\hat{z}$-component matches this axis in the standard frame (Fig.2B) by rotating the frame about $\hat{k} \times \hat{z}$, represented by the green arrows (along $\hat{x}$ in Fig.2A). Figs.2C and 2D present the case where the direction of $\hat{z}$-component of wave vector is opposite to the quantization axis. Regardless of the direction of the $\hat{z}$-component, the polarization angle (with respect to the $\hat{z}$-axis) is consistent in both cases as shown in Fig.2E. However, when the system is rotated by spatial rotation or boost of the frame, $\hat{z}$-component direction of the wave vector, resulting in varied polarization angles in the standard frame. As an example, under a frame rotation about $\hat{y}$-axis, Figs. 2F-J show the polarization angle varies in the standard frame depending on the direction of $\hat{z}$-component of wave vector but does not equate to the frame rotation angle due to the non-commutativity. A general observation as depicted in Fig.2J, is that polarization angle in the standard frame depends on the relative direction of $\hat{z}$-component of wave vector compared to the quantization axis (here the $\hat{z}$-axis), as shown in Fig.2J, which could lead to asymmetry in WRA. Here, $\hat{k}'$ and $\hat{p}'$ are represent the wave vector and polarization vector after the Lorentz transformation of $\hat{k}$ and $\hat{p}$, respectively.

### C. Violation of local-time reversal symmetry in WRA

In Fig.3, we illustrate scenarios for the study of local time reversal symmetry of the WRA, where photons are transmitted between satellites in polar orbits. The red shapes represent the wave vector $k$ and 4-velocity of each satellite $e_{\hat{i}}$, while blue shapes indicate those of the corresponding time-reversed case (Fig.3A), and the scenario with photon transmission in the opposite direction $-k^{\phi}$ in the azimuthal plane (Fig.3B). When dealing with local-time reversal symmetry, the special relativity and equivalence principle govern the behavior of the local wave vectors, represented as $\hat{\mathbf{k}}$. Under local-time reversal operator, special relativity dictates that the spatial components of the momentum of the particles reverses direction, while the energy (time component) remains unchanged. In accordance with the equivalence principle, this leads to sign flip of the local spatial components of momentum as shown in Fig.3A. Also, by the definition of the local-time reversal operator, the sign of timelike component of the tetrad fields should change because time reversal symmetry corresponds to reversing the direction of time, while keeping spatial directions unchanged. These conditions can be achieved by flipping the signs of affine parameter $\xi \to \xi' = -\xi$ and the proper time $\tau \to \tau' = -\tau$ such that

$$\left(e^{i}{}_{\mu}, e^{i}{}_{\mu}\right) = \left(\left(\mathrm{d}x^{\nu}/d\tau\right)g_{\mu\nu}, e_{i}{}^{\nu}g_{\mu\nu}\right) \underset{\substack{\xi \to \xi' \\ \tau \to \tau'}}{\longrightarrow} \left(\left(\mathrm{d}x^{\nu}/d\tau'\right)g_{\mu\nu}, e_{i}{}^{\nu}g_{\mu\nu}\right) = \left(-\left(\mathrm{d}x^{\nu}/d\tau\right)g_{\mu\nu}, e_{i}{}^{\nu}g_{\mu\nu}\right) = \left(-e^{i}{}_{\mu}, e^{i}{}_{\mu}\right),$$

$$\left(\left|\hat{\mathbf{k}}\right|, \hat{\mathbf{k}}\right) = \left(\left|\hat{\mathbf{k}}\right|, \left(\mathrm{d}x^{\mu}/\mathrm{d}\xi\right)e^{i}{}_{\mu}\right) \underset{\substack{\xi \to \xi' \\ \tau \to \tau'}}{\longrightarrow} \left(\left|\hat{\mathbf{k}}\right|, \left(\mathrm{d}x^{\mu}/\mathrm{d}\xi'\right)e^{i}{}_{\mu}\right) = \left(\left|\hat{\mathbf{k}}\right|, -\left(\mathrm{d}x^{\mu}/\mathrm{d}\xi\right)e^{i}{}_{\mu}\right) = \left(\left|\hat{\mathbf{k}}\right|, -\hat{\mathbf{k}}\right),$$

$$\nabla_{k} = \left(\mathrm{d}x^{\mu}/\mathrm{d}\xi\right)\nabla_{\mu} \underset{\substack{\xi \to \xi' \\ \tau \to \tau'}}{\longrightarrow} \left(\mathrm{d}x^{\mu}/\mathrm{d}\xi'\right)\nabla_{\mu} = \left(-\mathrm{d}x^{\mu}/\mathrm{d}\xi\right)\nabla_{\mu} = -\nabla_{k},$$

$$(6)$$

with $\hat{i} = 1$, $2$, and $3$. Under these transformations, the signs of local infinitesimal rotations, $\lambda^{\hat{i}}_{\ \hat{j}} = (\nabla_k e^{\hat{i}\mu})e_{\mu\hat{j}}$[37,42], are flipped while those of local infinitesimal boosts, $\lambda^{\hat{i}}_{\ \hat{j}} = (\nabla_k e^{\hat{i}\mu})e_{\mu\hat{j}} = (\nabla_k dx^\mu / d\tau)e_{\mu\hat{j}}$[37,42], remain unchanged as shown in Fig3C. The corresponding time reversed infinitesimal Wigner rotation angle rate becomes

$$\frac{d\psi_{\hat{k}}}{d\xi} \xrightarrow{\ \hat{\mathscr{T}}\ } \frac{d\psi_{-\hat{k}}}{d\xi'} \left( = -\frac{d\psi_{-\hat{k}}}{d\xi} \right) = -\lambda^{\hat{1}}_{\ \hat{2}} + \left[ \frac{-n^{\hat{1}}}{1-n^{\hat{3}}}\left( -\lambda^{\hat{0}}_{\ \hat{2}} - \lambda^{\hat{2}}_{\ \hat{3}} \right) + \frac{-n^{\hat{2}}}{1-n^{\hat{3}}}\left( \lambda^{\hat{0}}_{\ \hat{1}} - \lambda^{\hat{3}}_{\ \hat{1}} \right) \right], \tag{7}$$

(compare with Eq. (4)). Thus, the time-reversed total Wigner rotation angle can be written in terms of the infinitesimal local Lorentz transformation $\lambda^{\hat{a}}_{\ \hat{b}}$ and unit vector $n^{\hat{i}}$ along the direction of photon's momentum $\hat{\mathbf{k}}$ as

$$\begin{aligned} \psi_{\hat{k}}(\xi) = \int_0^\xi \frac{d\psi_{\hat{k}}}{d\xi} d\xi \quad &\xrightarrow{\ \hat{\mathscr{T}}\ } \quad \int_0^{-\xi} \left( \frac{d\psi_{-\hat{k}}}{d\xi'} \right) d\xi' = -\int_0^\xi \left( -\frac{d\psi_{-\hat{k}}}{d\xi} \right) d\xi \\ &= \int_0^\xi \lambda^{\hat{1}}_{\ \hat{2}} d\xi - \int_0^\xi \left[ \frac{n^{\hat{1}}}{1-n^{\hat{3}}}\left( \lambda^{\hat{0}}_{\ \hat{2}} + \lambda^{\hat{2}}_{\ \hat{3}} \right) + \frac{n^{\hat{2}}}{1-n^{\hat{3}}}\left( -\lambda^{\hat{0}}_{\ \hat{1}} + \lambda^{\hat{3}}_{\ \hat{1}} \right) \right] d\xi \end{aligned}. \tag{8}$$

Then, the WRA difference between a path of photons and the time-reversed one represented with red and blue arrows in Fig3A, respectively, is non-zero in general such that

$$\begin{aligned} \Delta\psi = \int_{\text{photon's path}} d\psi_{\hat{k}} - \hat{\mathscr{T}} \left( \int_{\text{photon's path}} d\psi_{\hat{k}} \right) &= \int_0^\xi \left( \frac{d\psi_{\hat{k}}}{d\xi} \right) d\xi + \int_0^\xi \left( -\frac{d\psi_{-\hat{k}}}{d\xi} \right) d\xi = \\ &= \int_{x^{\hat{a}}(0)}^{x^{\hat{a}}(\xi)} \partial_{\hat{a}} \psi_{\hat{k}} dx^{\hat{a}} - \int_{x^{\hat{a}}(0)}^{x^{\hat{a}}(\xi)} \left( \partial_{\hat{a}} \psi_{-\hat{k}} \right) dx^{\hat{a}} \\ &= \int_0^\xi \left[ \frac{n^{\hat{1}}\lambda^{\hat{2}}_{\ \hat{3}}}{1-\left(n^{\hat{3}}\right)^2} + \frac{n^{\hat{2}}\lambda^{\hat{3}}_{\ \hat{1}}}{1-\left(n^{\hat{3}}\right)^2} \right] d\xi + \int_0^\xi \left[ \frac{2n^{\hat{1}}n^{\hat{3}}\lambda^{\hat{0}}_{\ \hat{2}}}{1-\left(n^{\hat{3}}\right)^2} - \frac{2n^{\hat{2}}n^{\hat{3}}\lambda^{\hat{0}}_{\ \hat{1}}}{1-\left(n^{\hat{3}}\right)^2} \right] d\xi. \end{aligned} \tag{9}$$

For further verification of local-time reversal symmetry breakdown, in the Supplementary Information (SI) , we derive the violation of time reversal symmetry in Euler-Lagrangian framework. In both Eq. (8) and the result of Euler-Lagrangian mechanics, the violation comes from the effect of the choice of quantization axis yielding non-zero components for $n^{\hat{1}}$ and $n^{\hat{2}}$, leading to non-zero WRA difference between the path and the time-revsersed. This symmetry breakdown is caused by the difference between Loretnz transformations $L(\hat{k})$ from $\hat{k}_{\text{std}}$ to $\hat{k}$ and $L\left(\mathcal{T}\hat{k}\right)$ from $\hat{k}_{\text{std}}$ to $\mathcal{T}\hat{k}$. Here, $\hat{k}_{\text{std}}$ and $\mathcal{T}\hat{k}$ represent the wave vector of a photon in standard and local-time reversed frame, respectively. Accordingly, when choosing the quantization axis such that the effects of these Lorentz transformations remain in the WRA, the WRA exhibits a localized breakdown in time-reversal symmetry. Accordingly, by selecting a quantization axis parallel to the spatial component of the photon's wave vector, which avoids retaining the effects of these Lorentz transformations within the WRA, we can ensure that quantum communication systems maintain stability and reliability of quantum states, undisturbed by this localized time-reversal symmetry breakdown.

It should be also highlighted that the local space-inversion operator not only reverses the direction of momentum but also alters the sign of helicity. Consequently, considering the non-reciprocity of the Wigner Rotation Angle (WRA) that maintains the helicity, we focus on the local time-reversal operator instead (See SI).

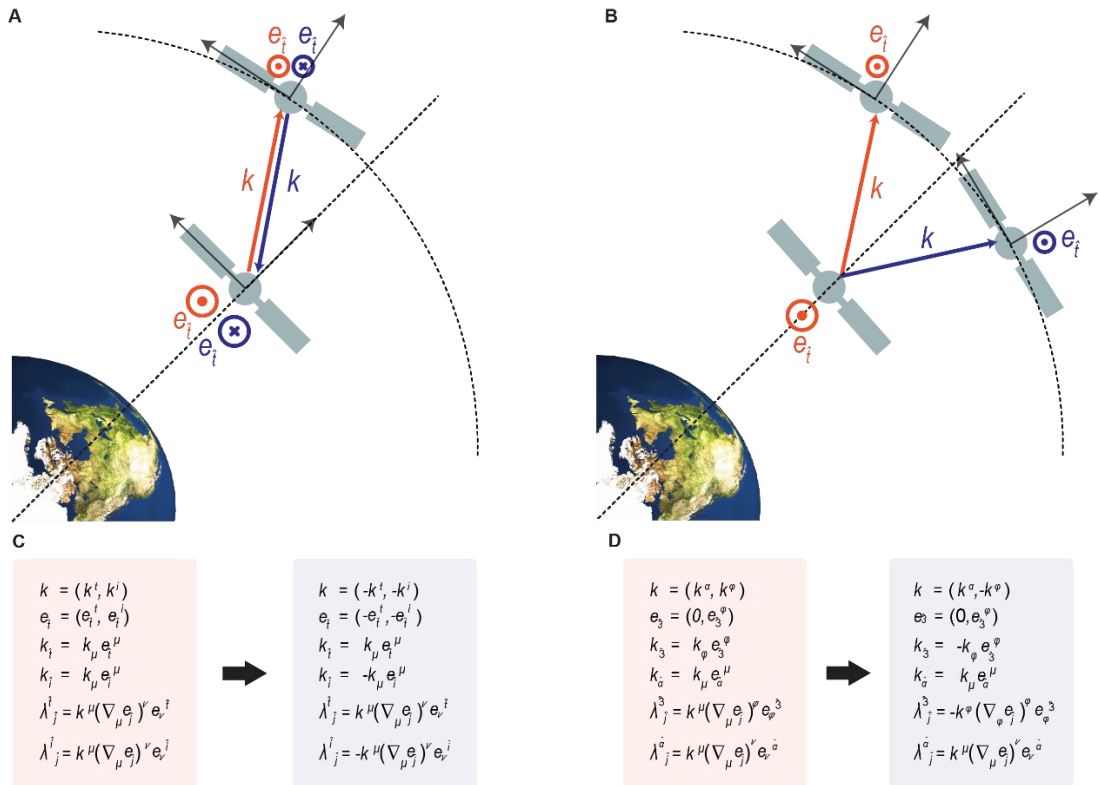

**Figure 3** Local time reversal symmetry violation and WRAs of wave vectors with opposite azimuthal component. Polar orbits of satellites are considered whose 4-velocity vectors are orthogonal to the equatorial plane as shown in Figs.3A and 3B. Under local-time reversal symmetry, photons sent between satellites with wave vector $k$ (depicted with the red arrow in Fig.3A), the signs of local spatial components should be flipped, while the local time-component (local frequency), remain unchanged, as dictated by the special relativity and equivalence

principle. The local-time reversed vectors are depicted with blue in Fig. 3A. Additionally, since by definition local-time reversal symmetry implies a sign flip of time component of tetrad $e_{\hat{i}}$, with no sign change in the same spatial components $e_{\hat{i}}$, the corresponding wave vector in Boyer-Lindquist coordinate bases, local boosts, and rotations should be transformed as shown in Fig.3C. When photons are sent with the opposite azimuthal wave vector component, (depicted by the blue arrow in Fig.3B), the corresponding transformation of the Christoffel symbols of the spherically symmetric Schwarzschild metric are given in Fig.3D, where $\hat{\alpha}$ is 0, 1, and 2. The original image of Earth used in Fig.3 is obtained from Yeongkwang Kim[47].

### D. Asymmetry in WRA under different sign of the momentum component along the quantization axis

For the case of an opposite azimuthal component of wave vector, the corresponding transformation of each component of wave vector, tetrads, and local Lorentz transformation is given in Fig.3d. In Schwarzschild spacetime ($J$=0), only the $\lambda^{\hat{3}}{}_{\hat{i}}$, with $\hat{i}$ =1 and 2, are dependent on the azimuthal component $k^{\phi}$ and is given as

$$-k^{\phi}\sqrt{1-\frac{r_s}{r}}F_{\hat{i}}\left(\sqrt{1-\frac{3r_s}{r}}\left(\theta-\frac{\pi}{2}\right)\right),\tag{10}$$

where $F_{\hat{i}}(x)=\begin{cases}\cos(x) \text{ for } \hat{i}=1\\ \sin(x) \text{ for } \hat{i}=2\end{cases}$. Accordingly, the WRA difference $\Delta\psi$ between two paths, represented with red (path 1) and blue arrows (path 2) in Fig.3b, is as follows:

$$\Delta\psi=\int_{\text{path1}}\mathrm{d}\psi-\int_{\text{path2}}\mathrm{d}\psi=\int_0^{\xi}\left[\frac{n^{\hat{1}}\lambda^{\hat{2}}{}_{\hat{3}}}{1-\left(n^{\hat{3}}\right)^2}+\frac{n^{\hat{2}}\lambda^{\hat{3}}{}_{\hat{1}}}{1-\left(n^{\hat{3}}\right)^2}\right]d\xi+\int_0^{\xi}\left[\frac{2n^{\hat{1}}n^{\hat{3}}\lambda^{\hat{0}}{}_{\hat{2}}}{1-\left(n^{\hat{3}}\right)^2}-\frac{2n^{\hat{2}}n^{\hat{3}}\lambda^{\hat{0}}{}_{\hat{1}}}{1-\left(n^{\hat{3}}\right)^2}\right]d\xi.\tag{11}$$

The relative WRA in Eq. (11) has the same form as that of the case of the local time-reversal symmetry violation in Eq. (9). Ascribed to the asymmetry illustrated in Fig.2 of the main text, the sign of $n^{\hat{3}}$ affects both scenarios depicted in Fig.3 through the denominator $1-\left(n^{\hat{3}}\right)^2$ of Eqs. (9) and (11). As the value of $n^{\hat{3}}$ increases, the denominator $1-\left(n^{\hat{3}}\right)^2$ decreases, and hence the value of the equations can be enhanced in the case of polar Earth orbits up to measurable orders. This also explains why observers in polar and equatorial orbits of Earth exhibit different WRA dependences based on the sign of azimuthal component wave vector $k^{\phi}$ of photons in the equatorial plane, where in the former case, the quantization axis (azimuthal direction) was set to be orthogonal to orbit plane, while for latter, the quantization axis was set to the polar (or zenith) direction. Since we set the photon trajectories to lie in the equatorial plane, the asymmetry induced by the form of the denominator in Eqs. (9) and (11) only affects the case of polar orbits. Moreover, by choosing the quantization axis along the wave vector, for which $n^{\hat{1}} = n^{\hat{2}} = 0$, the relative WRA of Eqs. (9) and (11) become zero, consistent with the result from the previous work[39]. This means the effect intertwined with local Lorentz transformations induces the asymmetry and violation of local-time reversal symmetry.

### E. Detecting WRA in an astronomical interferometer

The relative WRA induced by the effect from choice of the quantization axis can be measured near Earth in both quantum source and classical light source (Fig. 4). For the former, the combination of a Hong-Ou-Mandel (HOM) and Mach-Zehnder interferometer consisting of four satellites, each on polar orbits, can be utilized.; At Alice's local frame, two indistinguishable photon helicity states, $|1,1\rangle_{a_1,b_1}$, are launched, passing through a 50:50 beam splitter, after which a $\pi/2$-phase shift on one of the output ports $d_1$ is applied (Fig.4B). In this setup, we select a two-photon state with helicity $\sigma$ as an input state of the interferometer to observe the WRA $\psi$ in the coincidence rate. For the helicity state, WRA manifests as a phase shift of $\sigma\psi$ rather than a rotation angle of the polarization. This WRA-induced phase difference between the two paths could lead to the variation of the coincidence rate.

Using the creation operators of the two input $a_1^\dagger$ and $b_1^\dagger$ and two output ports $c_1^\dagger$ and $d_1^\dagger$ of the 50:50 beam splitter, the two-photon helicity states at each in input port can be written in terms of output-port operators as

$$
\begin{aligned}
|1,1\rangle_{a_1,b_1} &= a_1^\dagger b_1^\dagger |0,0\rangle \\
&= \frac{1}{2}\left(c_1^\dagger + d_1^\dagger\right)\left(c_1^\dagger - d_1^\dagger\right)|0,0\rangle = \frac{1}{2}c_1^{\dagger 2}|0,0\rangle - \frac{1}{2}d_1^{\dagger 2}|0,0\rangle. \\
&= \frac{1}{2}|2,0\rangle_{c_1,d_1} - \frac{1}{2}|0,2\rangle_{c_1,d_1}
\end{aligned}
\tag{12}
$$

Correspondingly, the state after the phase shift on an output port $d_1$ becomes $\left( |2,0\rangle_{c_1,d_1} - e^{i\pi/2} |0,2\rangle_{c_1,d_1} \right) / 2$.

Following this, the photons from each output port are sent from Alice to Bob to David, and from Alice to Charlie to David with $rk^\phi / k^r = \tan\alpha$ (as shown in Fig.4A), which is approximately the impact parameter ($b_{ph}$) of a photon's trajectory. Then, considering the relative Wigner rotation angle $\Delta\psi$ between the two arms of the interferometer, the photon states passing through output ports $c_{II}$ and $d_{II}$ of the second beam splitter (Fig.4C) are as follows:

$$
\begin{aligned}
&\frac{e^{\sigma i\Delta\psi}}{2} c_{I}^{\dagger 2} |0,0\rangle - i\frac{1}{2} d_{I}^{\dagger 2} |0,0\rangle \\
&= \frac{1}{4} e^{\sigma i\Delta\psi} \left( c_{II}^{\dagger 2} + 2c_{II}^{\dagger} d_{II}^{\dagger} + d_{II}^{\dagger 2} \right) |0,0\rangle - \frac{i}{4} \left( c_{II}^{\dagger 2} - 2c_{II}^{\dagger} d_{II}^{\dagger} + d_{II}^{\dagger 2} \right) |0,0\rangle \\
&= -i \left( \frac{ie^{\sigma i\Delta\psi} + 1}{4} \left( c_{II}^{\dagger 2} + d_{II}^{\dagger 2} \right) + \frac{-1 + ie^{\sigma i\Delta\psi}}{2} c_{II}^{\dagger} d_{II}^{\dagger} \right) |0,0\rangle \\
&= -i \left( \frac{ie^{\sigma i\Delta\psi} + 1}{2\sqrt{2}} \left( |2,0\rangle_{c_{II}d_{II}} + |0,2\rangle_{c_{II}d_{II}} \right) + \frac{-1 + ie^{\sigma i\Delta\psi}}{2} |1,1\rangle_{c_{II}d_{II}} \right),
\end{aligned}
\tag{13}
$$

where the relative WRA $\Delta\psi$ has the same form of Eq. (9). The relative WRA $\Delta\psi$ changes the linear combination of two indistinguishable photon states in Eq. (13), leading to the coincidence rate at David's frame, $(1 - \sin(\sigma\Delta\psi)) / 2$, depending on both the helicity and relative WRA.

In addition, we consider a Mach-Zehnder interferometer with a nearly monochromatic light source or a single-photon source. In Alice's local frame the classical light or a single photon enters through one of the input ports of the first beam splitter (as illustrated in the Fig.4D) and then in David's frame, the relative phase difference is measured after the light or the photon passes through the second beam splitter after aligning wave vectors to the quantization axis. By comparing the results from two different sources, it is possible to validate the equivalence between the Wigner rotation of photon states with the transformation of classical electric fields (proven in the Method) [34,37,42]. Moreover, this Mach-Zehnder interferometer could underscore the interplay of the WRA and non-local properties of a photon when a single-photon source is employed.

The relative WRA differences between the two arms of an interferometer near the Earth have a small order of $10^{-4}$ degrees compared to a few degrees (See Fig.S3D) for an interferometer around M87*. Considering the precision of LIGO and LISA where a few tens of attometer and picometer precision are required with near-infrared light respectively, $O(10^{-4})$ degrees of the relative WRA (shown in Fig.4E) should be measurable. Moreover, while the spinning angular momentum of M87* black hole would lead to a noticeable difference in the WRA, on the $O(10^{-2})$ degrees, but that of Earth does not significantly affect the WRA to measurable order for both equatorial and polar orbits, as supported by the results of the Gravity Probe B (See SI).

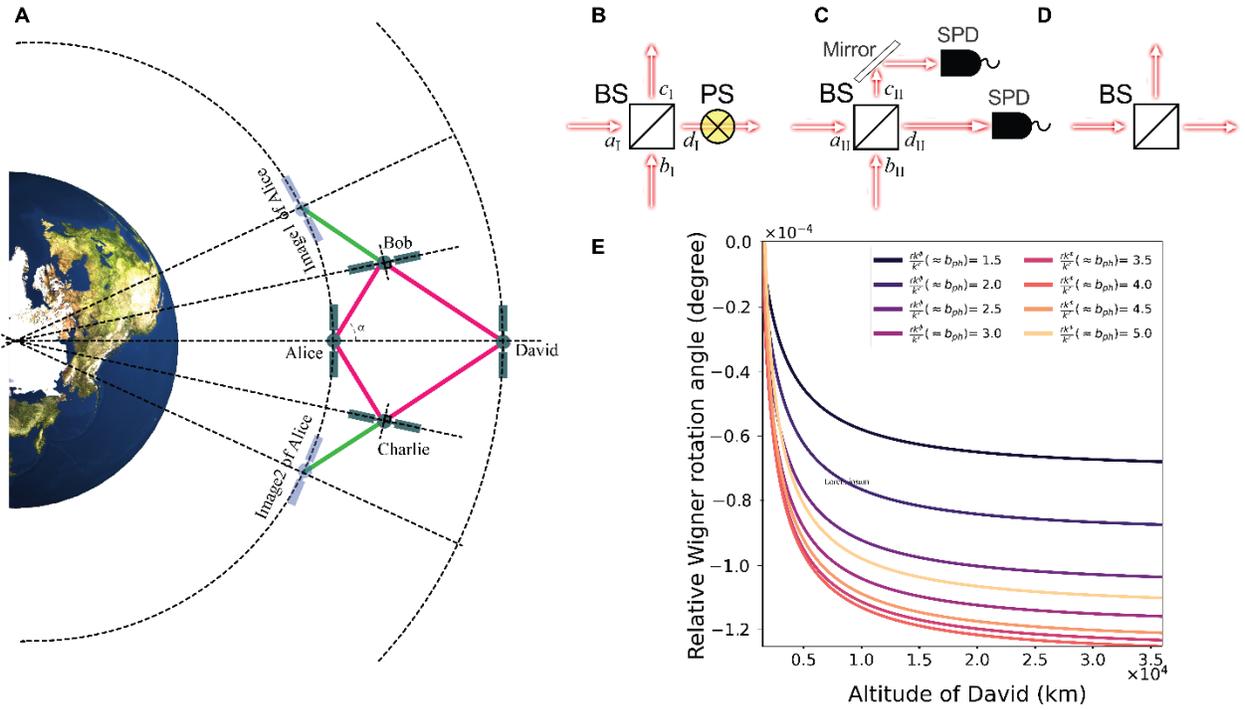

**Figure 4** Astronomical Mach-Zehnder interferometry setup designed to observe WRA (depicted in Fig.4A). For a quantum source scenario, two indistinguishable photons enter two input ports $a_I$ and $b_I$, respectively, of the first beam splitter (BS). In front of one output port $b_I$, $\pi/2$-phase shift is applied as shown in Fig.4B. In David's frame, photons are detected with single photon detectors (SPDs) after being aligned along the quantization axis depicted in Fig.4C. With Hong-Ou-Mandel (HOM) effect, photons emit from each output port with ½ possibility with $rk^\phi/k^r = \tan\alpha$. For a classical source one, light passes through as in Fig.4D and is sent to David with $rk^\phi/k^r = \tan\alpha$. The relative phase difference is measured in David's frame, assuming that Allice is at the altitude of 300 km, as shown in Fig.4E. WRA from Bob and Charlie to David are calculated as integrating infinitesimal WRA along photon's geodesics from image1 and 2 of Alice to David and subtract the part integrated along images to Bob and Charlie depicted with green lines. The original image of Earth used in Fig.4 is obtained from Yeongkwang Kim[47].

III DISCUSSION

Polarization measurements of photons offer a room a window to unveil a variety of phenomena, ranging from the quantum gravity[26] to the possibility of violations of the Einstein Equivalence Principle[27]. The polarization of a photon undergoes alterations as it interacts with the curvature of spacetime during its passage. In this work, we uncover a new facet: the non-reciprocity that emerges from the strategic choice of a quantization axis. Numerical calculations show that in different environments, near-Earth to near a black hole, the WRA can be measured as ten times larger than the rotation angle of the frame by exploiting this non-reciprocity (See SI). This finding suggests that the non-reciprocity could be harnessed to detect subtle polarization rotations under varying gravitational conditions.

This non-reciprocity extends its influence beyond the usual scope of the Wigner rotation angle (WRA), associated with frame variations in spacetime. When a standard frame is established with the chosen quantization axis, there would be a sequence of three Lorentz transformations including a polarization rotation, regardless of its origins: transforming the standard frame to the original, applying the polarization rotation around the photon's wave vector, and returning to the standard frame. While in this work we only consider polarization rotations in classical gravitational fields, following the Einstein Equivalence Principle, polarization rotation can be induced by various factors such as quantum gravity and violations of the Einstein Equivalence Principle—If the polarization rotations are considered as an SO(2) rotation on the polarization plane in each formalism, we could still utilize the non-reciprocity by introducing a standard frame defined with a tailored direction of the quantization axis.

It has been shown that specific violations of the Einstein Equivalence Principle (EEP) would lead to a rotation in the plane of polarization of radiation from distant radio sources[28]. In the effective field theory approach within quantum gravity, Lorentz- and CPT-violating dispersion relations can arise, leading to different group velocities for different polarization states[26]. This results in a polarization rotation during propagation, as described by a modified dispersion relation. These additional rotations in the polarization plane can be examined using the standard frame with a tailored direction of the quantization axis, thus allowing us to utilize the non-reciprocity. Moreover, validating the non-reciprocity induced from the choice of quantization axis could be feasible on an optical table. The non-reciprocity itself can also arise from inherent non-commutativity of rotations with no need for a boost as depicted in Fig. 2, which can be effectively realized using polarizers and mirrors by effectively simulating rotations. Thus, optical table experiments could demonstrate and validate the non-reciprocity in a controlled laboratory environment.

## Data availability

The data supporting the plots and other findings presented in this study are openly accessible on our GitHub repository at the following link: https://github.com/gksshcks/WRA_For_different_impact_parameters.

## Code availability

The Mathematica codes that support the data used in this paper and other finding are available from the corresponding author upon reasonable request.

## Method

All the data in the GitHub link have been obtained using a Mathematica notebook (Wolfram Research Europe Ltd, Long Hanborough, UK).

Geodesics in Kerr spacetime

The Kerr metric is applied to model spacetimes considered in this work with spin angular momentum ($J/Mc = a$) of a gravitating object, which is given by[48]

$$ds^2 = -\left(1 - \frac{r_s r}{\Sigma}\right)c^2 dt^2 + \frac{\Sigma}{\Delta}dr^2 + \Sigma d\theta^2 + \left(r^2 + a^2 + \frac{r_s r a^2}{\Sigma}\sin^2\theta\right)\sin^2\theta d\phi^2 - \frac{2r_s r a\sin^2\theta}{\Sigma}c dt d\phi, \tag{14}$$

where $\Sigma \equiv r^2 + a^2 cos^2\theta$, $\Delta \equiv r^2 ? r_s r + a^2$, and $r_s \equiv 2GM/c^2$ is the Schwarzschild radius. $G$ and $c$ are the gravitational constant and the speed of light. $J$ and $M$ are the angular momentum and the mass of the gravitating object, respectively.

The geodesics of a particle in Kerr spacetime is described by[48]

$$\begin{pmatrix} \dfrac{dt}{d\xi} \\[2mm] \dfrac{dr}{d\xi} \\[2mm] \dfrac{d\theta}{d\xi} \\[2mm] \dfrac{d\phi}{d\xi} \end{pmatrix} = \begin{pmatrix} \dfrac{1}{\Delta\Sigma}\left(E\left[\left(r^2 + a^2\right)^2 - \Delta a^2\sin^2\theta\right] - 2Mra\Phi\right) \\[2mm] \pm\dfrac{1}{\Sigma}\sqrt{\left(E\left(r^2 + a^2\right) - a\Phi\right)^2 - \Delta\left(K + \delta_1 r^2\right)} \\[2mm] \pm\dfrac{1}{\Sigma}\sqrt{K - \delta_1 a^2\cos^2\theta - \left(aE\sin\theta - \Phi/\sin\theta\right)^2} \\[2mm] \dfrac{1}{\Delta\Sigma}\left(2MraE + (\Sigma - 2Mr)\Phi/\sin^2\theta\right) \end{pmatrix}. \tag{15}$$

Here, $K = C + (\Phi - aE)^2$. The parameters ($\xi$, $\delta_1$) are (affine parameter, 0) or (proper time, 1) for null or time-like geodesics, respectively. $E$, $\Phi$, and $C$ are the conserved quantities, corresponding to energy at infinity, axial angular momentum of an orbiting particle, and Carter's (fourth) constant, defined as $C = \left(d\theta/d\xi\right)^2 + \cos^2\theta\left(a^2\left(\delta_1^2 - E^2\right) + \left(\Phi cosec\theta\right)^2\right)$[48-50], respectively.

<u>Co-moving tetrads with a satellite, parallel transported along its geodesic</u>

To describe local inertial frames of satellites, we obtain co-moving and non-spinning tetrads $\{e_{\hat{i}} = (e_{\hat{i}}{}^{t}, e_{\hat{i}}{}^{r}, e_{\hat{i}}{}^{\theta}, e_{\hat{i}}{}^{\phi}) | \hat{i} = 0,1,2,$ and $3\}$ by transforming Marck's $\{\lambda_{\hat{i}} = (\lambda_{\hat{i}}{}^{(0)}, \lambda_{\hat{i}}{}^{(1)}, \lambda_{\hat{i}}{}^{(2)}, \lambda_{\hat{i}}{}^{(3)}) | \hat{i} = 0,1,2,$ and $3\}$ [48] from Carter's symmetric bases $(\omega^{(0)}, \omega^{(1)}, \omega^{(2)}, \omega^{(3)})$ back to the Boyer-Lindquist coordinate ones $(dt, dr, d\theta, d\phi)$. Co-moving condition ensures that the reference frame moves along with the satellite, providing a local rest frame for measurements, and non-spinning condition is essential to eliminate any arbitrary rotational effects within the reference frame. The Marck's solution satisfies the two conditions by obtaining the tetrads parallel-transported along a test particle's geodesics [49], specifically those of satellites in our case.

In details, the Marck's tetrads are defined as follows [49]:

$$
\begin{pmatrix} \lambda_0 \\ \lambda_1 \\ \lambda_2 \\ \lambda_3 \end{pmatrix} = \begin{pmatrix} 1 & 0 & 0 & 0 \\ 0 & \cos\Psi & 0 & -\sin\Psi \\ 0 & 0 & 1 & 0 \\ 0 & \sin\Psi & 0 & \cos\Psi \end{pmatrix} \begin{pmatrix} \sqrt{\Delta/\Sigma}\left(e_0{}^t - a\sin^2\theta\, e_0{}^\phi\right) & \sqrt{\Sigma/\Delta}e_0{}^r & \sqrt{\Sigma}e_0{}^\theta & \left(ae_0{}^t - (r^2+a^2)e_0{}^\phi\right)\sin\theta\, e_0{}^t/\sqrt{\Sigma} \\ \alpha\sqrt{\Sigma/K\Delta}re_0{}^r & \alpha\sqrt{1/K\Sigma}\Sigma r\left\{E(r^2+a^2) - a\Phi\right\} & \beta\sqrt{1/K\Sigma}a\cos\theta(aE\sin\theta - \Phi/\sin\theta) & -\beta\sqrt{\Sigma/K}a\cos\theta e_0{}^\theta \\ \sqrt{\Sigma/K}a\cos\theta e_0{}^r & \sqrt{1/K\Sigma}a\cos\theta\left\{E(r^2+a^2) - a\Phi\right\} & -\sqrt{1/K\Sigma}r(aE\sin\theta - \Phi/\sin\theta) & \sqrt{\Sigma/K}re_0{}^\theta \\ \alpha\sqrt{1/\Sigma\Delta}\left\{E(r^2+a^2) - a\Phi\right\} & \sqrt{1/K\Sigma}a\cos\theta\left\{E(r^2+a^2) - a\Phi\right\} & \beta\sqrt{\Sigma}e_0{}^\theta & \beta\sqrt{1/\Sigma}\left(aE\sin\theta - \Phi/\sin\theta\right) \end{pmatrix}
\tag{16}
$$

The parameter $\Psi$ is introduced for the parallel-transport condition of spacelike component of tetrads along the geodesics of observers, i.e., satellites. The parameter $\psi$ can be obtained by integrating the following equation [49],

$$
\frac{d\Psi}{d\xi} = \frac{K^{1/2}}{\Sigma}\left(\frac{E\left(r^2+a^2\right) - a\Phi}{r^2+K} + a\frac{\left(\Phi - aE\sin^2\theta\right)}{K - a^2\cos^2\theta}\right).
\tag{17}
$$

we solve the Eq. (17) with respect to $(r, \theta, \phi)$ from (the radius of Earth, $\pi/2, -\pi$) to (an altitude of 36000 km, $\pi$, $\pi$) with the intervals (100km, $\pi/10$, $\pi/20$) for the case of near-Earth and for the black hole, we solve the equation from $(4.5r_s, \pi/2, -\pi)$ to $(4.5r_s + 2\times10^{11}$ km, $\pi$, $\pi)$ with the intervals $(2\times10^9$ km, $\pi/10$, $\pi/20)$.

For equatorial orbits, as $r$ and $\theta$ should be constant, Eq. (17) can be rewritten as

$$
\frac{d\Psi}{d\phi} = \frac{K^{1/2}}{\Sigma}\left(\frac{E\left(r^2+a^2\right) - a\Phi}{r^2+K} + a\frac{\left(\Phi - aE\sin^2\theta\right)}{K - a^2\cos^2\theta}\right)\frac{d\xi}{d\phi}.
\tag{18}
$$

In the case of polar orbits, if $\Psi$ depends on $r$, $\theta$, and $\phi$, $d\Psi/d\xi$ can be rewritten as $(\partial\Psi(r,\theta,\phi)/\partial r)(dr/d\xi) + (\partial\Psi(r,\theta,\phi)/\partial\phi)(d\phi/d\xi) + (\partial\Psi(r,\theta,\phi)/\partial\theta)(d\theta/d\xi)$ using the chain rule. Since the four-velocity vector is independent of $\phi$ (as shown in Eq. (15)), the second term $(\partial \Psi(r,\theta,\phi)/\partial \phi)(d\phi/d\xi)$ cannot be generally zero. Given that Eq. (17) are also independent of $\phi$, this leads to a contradiction. Therefore, we conclude $\Psi$ is independent of $\phi$. Accordingly, considering, as the case of equatorial orbit, $r$ is constant on the orbits, the following equation dictates the parameter $\psi$ of the tetrads parallel-tranported on polar orbits:

$$\frac{d\Psi}{d\theta} = \frac{K^{1/2}}{\Sigma}\left(\frac{E(r^2+a^2)-a\Phi}{r^2+K}+a\frac{(\Phi-aE\sin^2\theta)}{K-a^2\cos^2\theta}\right)\frac{d\xi}{d\theta}. \tag{19}$$

Then, transforming the Marck's from Carter's symmetric bases to the Boyer-Lindquist coordinate ones, we obtain the tetrads $e_i$ used in this paper such that

$$\begin{pmatrix} e_{\hat{i}}^{\ t} \\ e_{\hat{i}}^{\ r} \\ e_{\hat{i}}^{\ \theta} \\ e_{\hat{i}}^{\ \phi} \end{pmatrix} = \begin{pmatrix} \dfrac{a^2+r^2}{\sqrt{\Delta\Sigma}} & 0 & 0 & -\dfrac{a\sin\theta}{\sqrt{\Sigma}} \\ 0 & \sqrt{\dfrac{\Delta}{\Sigma}} & 0 & 0 \\ \dfrac{a}{\sqrt{\Delta\Sigma}} & 0 & 0 & -\dfrac{\csc\theta}{\sqrt{\Sigma}} \\ 0 & 0 & -\sqrt{\dfrac{1}{\Sigma}} & 0 \end{pmatrix} \begin{pmatrix} \lambda_{\hat{i}}^{\ (0)} \\ \lambda_{\hat{i}}^{\ (1)} \\ \lambda_{\hat{i}}^{\ (2)} \\ \lambda_{\hat{i}}^{\ (3)} \end{pmatrix}. \tag{20}$$

The tetrads $(e_{\hat{t}}, e_{\hat{r}}, e_{\hat{\theta}}, e_{\hat{\phi}})$ obtained from Eq. (20) are asymptotically parallel to the unit vectors of global coordinates $(\partial_t, \partial_r, \partial_\theta, \partial_\phi)$ as $r$ goes to infinity. This transformation also ensures the tetrads obtained from Marck's become equivalent to those used to study of WRA in Schwarzschild spacetime in our previous work[37] by setting spin angular momentum ($J/Mc = a$) of gravitating object as zero.

Also, we additionally rotate the tetrads about the local first axis to choose the local third axis as that of geodetic precession, compensating the frame-dragging effect. The corresponding rotation angle $\Xi$ is obtained by

$$-\cos\Xi e_2^{\ \theta} + \sin\Xi e_3^{\ \theta} = 0. \tag{21}$$

This approach is used to ensure a consistent comparison with our previous work[37] on the Schwarzschild metric, In that previous work, the local third axis, the quantization axis, was chosen to be the geodetic precession axis to isolate the geodetic precession.

## Wigner Rotation and the Complementarity of Classical and Quantum Theories: Equivalence between Wigner rotation of a photon and classical polarization rotation

The polarization vector, denoted as $\varepsilon(\hat{\mathbf{p}}, \sigma=\pm 1)$, is defined with $R(\hat{\mathbf{p}})$ rotating the wave vector $\hat{\mathbf{k}}_{std}=(1,0,0,1)$ in the standard frame into the direction of arbitrary momentum $\hat{\mathbf{p}}$ such that[32]

$$\varepsilon\left(\hat{\mathbf{p}}, \sigma=\pm 1\right) = \frac{R(\hat{\mathbf{p}})}{\sqrt{2}} \varepsilon\left(\hat{\mathbf{k}}_{std}, \sigma=\pm 1\right) = \frac{R(\hat{\mathbf{p}})}{\sqrt{2}}\begin{pmatrix} 0 \\ 1 \\ \pm i \\ 0 \end{pmatrix}. \tag{22}$$

Here, $\sigma$ represents the helicity. Throughout this paper, the hatted letter represents local or global flat spacetime. If the polarization vector is treated as a four-vector, it can be directly subjected to a Lorentz transformation $\Lambda$ and be rewritten as follows:

$$\begin{aligned}
\Lambda \varepsilon_{\sigma=\pm 1}(\hat{\mathbf{p}}) &= L(\Lambda\hat{\mathbf{p}})\left\{L^{-1}(\Lambda\hat{\mathbf{p}})\Lambda L(\hat{\mathbf{p}})\right\}L^{-1}(\hat{\mathbf{p}})\varepsilon(\hat{\mathbf{p}},\sigma=\pm 1) \\
&= L(\Lambda\hat{\mathbf{p}})W(\Lambda,\hat{\mathbf{p}})\varepsilon(\hat{\mathbf{k}}_{std},\sigma=\pm 1) \\
&= L(\Lambda\hat{\mathbf{p}})S(\alpha,\beta)R_z\left(\psi(\Lambda,\hat{\mathbf{p}}/|\hat{\mathbf{p}}|)\right)\varepsilon(\hat{\mathbf{k}}_{std},\sigma=\pm 1) \\
&= L(\Lambda\hat{\mathbf{p}})\left\{e^{i\sigma\psi(\Lambda,\hat{\mathbf{p}}/|\hat{\mathbf{p}}|)}\varepsilon(\hat{\mathbf{k}}_{std},\sigma=\pm 1) + \frac{\alpha+i\beta}{\sqrt{2}\left|\hat{\mathbf{k}}_{std}\right|}\hat{\mathbf{k}}_{std}\right\} \\
&= \left\{R(\Lambda\hat{\mathbf{p}})e^{i\sigma\psi(\Lambda,\hat{\mathbf{p}}/|\hat{\mathbf{p}}|)}\varepsilon(\hat{\mathbf{k}}_{std},\sigma=\pm 1) + \frac{\alpha+i\beta}{\sqrt{2}\left|\hat{\mathbf{k}}_{std}\right|}L(\Lambda\hat{\mathbf{p}})\hat{\mathbf{k}}_{std}\right\} \\
&= \left\{e^{i\sigma\psi(\Lambda,\hat{\mathbf{p}}/|\hat{\mathbf{p}}|)}\varepsilon(\Lambda\hat{\mathbf{p}},\sigma=\pm 1) + \frac{\alpha+i\beta}{\sqrt{2}\left|\hat{\mathbf{k}}_{std}\right|}\Lambda\hat{\mathbf{p}}\right\}
\end{aligned} \tag{23}$$

with $\quad S(\alpha,\beta)=\begin{pmatrix} 1+\dfrac{\alpha^2+\beta^2}{2} & \alpha & \beta & -\dfrac{\alpha^2+\beta^2}{2} \\ \alpha & 1 & 0 & -\alpha \\ \beta & 0 & 1 & -\beta \\ \dfrac{\alpha^2+\beta^2}{2} & \alpha & \beta & 1-\dfrac{\alpha^2+\beta^2}{2} \end{pmatrix}$ [32]. The Wigner's little group $W(\Lambda,\hat{\mathbf{p}})$, defined as $L^-$

$^1(\Lambda\hat{\mathbf{p}})\Lambda L(\Lambda)$, which is a subgroup of the Lorentz group, leaves the wave vector in the standard frame invariant and

can be decomposed into $S(\alpha,\beta)R_z(\psi(\Lambda,\ \hat{\mathbf{p}}/|\hat{\mathbf{p}}|))$[32] where $S(\alpha,\beta)$ is a subgroup isomorphic to the translation of a Euclidean plane and $R_z(\psi)$ represents the rotation about $z$-axis by $\psi$. The rotation angle $\psi(\Lambda,\ \hat{\mathbf{p}}/|\hat{\mathbf{p}}|)$ corresponds to the Wigner rotation angle. Here, we use the decomposition of the Lorentz transformation $L(\hat{\mathbf{p}})$, which maps the wave vector $\hat{\mathbf{k}}_{\text{std}}$ in the standard frame to a wave vector $\hat{\mathbf{p}}$, into $R(\hat{\mathbf{p}})B_z(|\hat{\mathbf{p}}|)$ along with the invariance of the polarization vector under a boost $B_z(|\hat{\mathbf{p}}|)$ in the z-direction in the standard frame[51]. As the polarization vector is essentially a three-dimensional vector in the spatial part of the spacetime, the corresponding representation of Lorentz transformation $U(\Lambda)$ can be defined as

$$U\left(\Lambda\right)\varepsilon\left(\Lambda\hat{\mathbf{p}},\sigma=\pm1\right)=\Lambda\varepsilon\left(\hat{\mathbf{p}},\sigma=\pm1\right)-\frac{\left(\Lambda\varepsilon\left(\hat{\mathbf{p}},\sigma=\pm1\right)\right)^{\hat{0}}}{\left(\Lambda\hat{\mathbf{p}}\right)^{\hat{0}}}\Lambda\hat{\mathbf{p}}=e^{i\sigma\psi\left(\Lambda,\hat{\mathbf{p}}/|\hat{\mathbf{p}}|\right)}\varepsilon\left(\Lambda\hat{\mathbf{p}},\sigma=\pm1\right). \quad (24)$$

Here, it is used that the time component of Lorentz transformed $\left(\Lambda\varepsilon\right)^{\hat{0}}$ corresponds to $\frac{+i}{\sqrt{2}\left|\hat{\mathbf{k}}_{std}\right|}\left(\Lambda\hat{\mathbf{p}}\right)^{\hat{0}}$, since polarization vector $\varepsilon$ has no time component.

In classical description, if the light can be described as monochromatic and circular polarized wave and then its corresponding potential four vector $\Phi$ is the function of inner product between momentum $\hat{\mathbf{p}}$ and position $\hat{x}$ — $\Phi=\varepsilon(\hat{\mathbf{p}},\sigma=\pm1)\phi(\hat{\mathbf{p}}\cdot\hat{x})$ — the electric fields have the same relation of Eq. (24) under the Lorentz transformation $\Lambda$. It implies that a Lorentz transformation of classical light, ensemble of photons, can have the Wigner rotation angle within the framework of Coulomb gauge fixing. Here, $\varepsilon(\hat{\mathbf{p}},\sigma=\pm1)$ is the four vector defined as Eq. (22) and $\phi(\hat{\mathbf{p}}\cdot\hat{x})$ is the scalar function. In details, the form of gauge-independent electromagnetic field derived from a Lorentz transformed potential field is as follows[34]:

$$\begin{aligned}E^{\hat{i}}\left(\hat{x}\right)&=-\left(\Lambda\varepsilon\left(\hat{\mathbf{p}},\sigma=\pm1\right)\right)^{\hat{i}}\partial^{\hat{0}}\phi\left(\hat{\mathbf{p}}\cdot\hat{x}\right)-\left(\Lambda\varepsilon\left(\hat{\mathbf{p}},\sigma=\pm1\right)\right)^{\hat{0}}\partial^{\hat{i}}\phi\left(\hat{\mathbf{p}}\cdot\hat{x}\right)\\&\propto\left(\Lambda\varepsilon\left(\hat{\mathbf{p}},\sigma=\pm1\right)\right)^{\hat{i}}\left(\Lambda\hat{\mathbf{p}}\right)^{\hat{0}}-\left(\Lambda\varepsilon\left(\hat{\mathbf{p}},\sigma=\pm1\right)\right)^{\hat{0}}\left(\Lambda\hat{\mathbf{p}}\right)^{\hat{i}}\\&\propto\left(\Lambda\varepsilon\left(\hat{\mathbf{p}},\sigma=\pm1\right)\right)^{\hat{i}}-\frac{\left(\Lambda\varepsilon\left(\hat{\mathbf{p}},\sigma=\pm1\right)\right)^{\hat{0}}}{\left(\Lambda\hat{\mathbf{p}}\right)^{\hat{0}}}\left(\Lambda\hat{\mathbf{p}}\right)^{\hat{i}}.\end{aligned} \quad (25)$$

where $E^{\hat{i}}(\hat{x})$ is $\hat{i}$-th component of electric vector field. It is worth mentioning, while the Wigner rotation can be described within the classical framework of electromagnetic waves, classical description cannot account for a super-positioned state of a single photon — a quantum mechanical phenomenon — but only the average effect of Lorentz transformations on a photon ensemble. In other words, classical description does not distinguish whether each photon itself becomes super-positioned or photons are statistically in one of two different spin states. Accordingly, the correspondence between classical and quantum descriptions of Wigner rotation of a photon showcases the deep connection between the two frameworks and further illustrates the complementary nature of classical and quantum theories in explaining physical phenomena within their respective domains of applicability rather than negating the other.

Furthermore, to derive the equivalent relation, it is imperative that the potential field of light be a four-vector, but not the electric or magnetic fields, and that the light exhibits circular polarization — the unique discretized form of polarization corresponding to the quantized helicity. This result underscores the primacy of the Wigner rotation description for the properties of the electric field under a Lorentz transformation since the Wigner rotation is rooted in intrinsic properties of the Wigner's little group depending on a particle's spin and the relative direction of a photon's path compared to the quantization axis of the spin[32,52]. This perspective could furnish a more foundational approach to understanding the dynamics of spacetime transformations.

This Wigner rotation on a quantum state should be differentiated from that in the context of Thomas precession observed in spinning particles or rotating macroscopic gyroscopes undergoing curvilinear motion[53,54]. When an observer in an accelerated frame of reference carries out a sequence of infinitesimal Lorentz transformations (boosts), an additional rotation manifests due to the non-commutativity of these transformations (a well-known special relativistic effect). Meanwhile, in the quantum domain, while Wigner rotation itself is induced from the non-commutativity of boosts as Thomas precession, its application to quantum states results in super-positioned spin states of a photon with different phases, called Wigner rotation angles, — distinguished from those which can be obtained from polarizer rotation. The Wigner rotation angle becomes observable via spin of particles, i.e., Wigner rotation induced from the little group does not affect spinless particles. Thus, the distinctiveness of the Wigner rotation angle is characterized by the introduction of superposition spin states without any rotation of the polarizer.

## Acknowledgements

This work was supported by Korean Ministry of Science and ICT through the National Information Agency (NIA). D.A. was also supported by Korea National Research Foundation (NRF) grant No. NRF-2023R1A2C1003570, ICT R&D program of MSIT/IITP 2021-0-01810, RS-2023-00225385, RS-2024-00422330, AFOSR grant FA2386-21-1-0089, and AFOSR grant FA2386-22-1-4052. P.M.A. would like to acknowledge support from the Air Force Office of Scientific Research (AFOSR). Any opinions, findings, and conclusions or recommendations expressed in this material are those of the author(s) and do not necessarily reflect the views of the Air Force Research Laboratory.

## Author contributions

H.N. introduced the HOM interferometer concept for measuring WRA differences, conducted numerical simulations, and developed a theoretical framework to demonstrate the non-reciprocity and violation of local-time reversal symmetry in WRA. P.M.A., W.M., and D.A. reviewed and validated the paper's arguments, suggesting an exploration of WRA in Kerr spacetime. All the authors contributed to the manuscript's composition.

## Competing interests

The authors declare no competing interests.

**Supplementary Information for Non-Reciprocity in Photon Polarization based on direction of polarizer under Gravitational Fields**

**Authors:** Hansol Noh[1,2], Paul M. Alsing[3], Warner A. Miller[3], and Doyeol Ahn[1,3*],

**Affiliations:**

[1]Department of Electrical and Computer Engineering and Center for Quantum Information Processing, University of Seoul, 163 Seoulsiripdae-ro, Dongdaemun-gu, Seoul, 02504, Republic of Korea

[2]Department of Electrical and Computer Engineering, Seoul National University Seoul 08826, Republic of Korea.

[3]Department of Physics, Florida Atlantic University, 777 Glades Road, Boca Rat Raton, FL 3431-0991, USA.

*Correspondence to: dahn@uos.ac.kr.

**Euler-Lagrange equation and local time-reversal symmetry breakdown of Wigner-rotated photon**

In this study, we employ non-spinning tetrads that are parallel transported along the geodesics of satellites (massive particles) orbiting around Earth. However, the Wigner rotation angle (WRA) should be calculated with covariant derivatives of tetrads along the geodesics of photons (massless particles). Consequently, while the variation of WRAs can be measured by comparing the phases of photons observed in local frames at initial and final events, we can alternatively introduce a fictitious observer whose local frame is described with a set of chosen tetrads defined along events on null geodesics. In the latter case, considering the acceleration of frames can be interpreted as the sequential applications of different boosts over time, the application of different local Lorentz transformations induced from tetrad variations along null geodesics leads to fictitious accelerations. Accordingly, as in the case of the Coriolis force, the fictitious acceleration can break time reversal symmetry in the local non-inertial frame.

We demonstrate the breakdown of time reversal symmetry by showing that time-reversed photon state does not follow the Euler-Lagrange equation obtained. To achieve this, we first approximate the spacetime of Earth as being geodesically complete, i.e., the paths of freely falling particles can be extended near the Earth indefinitely without encountering singularities or other obstructions by taking advantages of much smaller Schwarzschild radius of Earth ( $\sim$ 9mm) than the size of system. Considering the variations of local inertial frames defined with tetrad fields $e^{\hat{a}}{}_{\mu}(x)$ are mapped into others by local Lorentz transformations, we can introduce Lorentz-invariant four-dimensional volume of each infinitesimal local tangent space along the geodesics depicted with black boxes in Fig.S1. By continuously placing each of infinitesimal box, we may consider an "extended" flat spacetime $\{\hat{\mathbf{x}}_{\mathbf{B}}\}$, which is not "flat spacetime" with zero curvature but rather it is an approximated description of spacetime by assuming a smooth integration of infinitesimal boxes. Moreover, as the boundary of the neighborhood can be defined with parallel transportation of tetrad fields at $\hat{\mathbf{x}}_{\mathbf{B}} = 0$, there may be no boundary deficit in covering the entire spacetime. A similar approach has been reported to address the validity of Fourier transformation on any non-compact geodesically complete manifold[55].

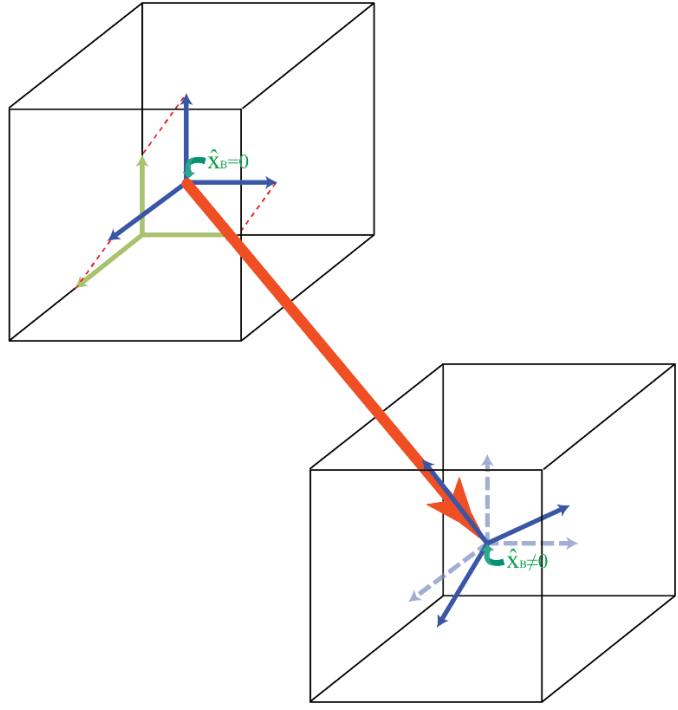

**Fig. S1** An infinitesimal box is parallel transported from $\hat{x}_B = 0$ along the geodesics depicted with a red arrow. The boundaries of the boxes are defined with tetrad fields, represented by blue arrows, parallel transported into light green arrows along red dot line representing geodesics of photons. The box is parallel transported along other photon geodesics. In each box, local frame is defined with tetrad fields which are Lorentz transformed from the original tetrad fields defined at $\hat{x}_B = 0$.

In this system, Einstein's equivalence principle supports that a single-component vector potential $A_i(\hat{x})\left(=\phi(\hat{x})\right)$ follows the Klein-Gordon equation. The corresponding field has the form:

$$A_i(\hat{x}) = \phi_{\text{scalar}}(\hat{x}) = \int \mathrm{d}^3\hat{\mathbf{k}} \, \frac{1}{\sqrt{(2\pi)^3 2k^{\hat{0}}}} \left( a_i(\hat{\mathbf{k}}) e^{-i\left(k_{\hat{\mu}} x^{\hat{\mu}}\right)} + a_i^\dagger(\hat{\mathbf{k}}) e^{i\left(k_{\hat{\mu}} x^{\hat{\mu}}\right)} \right) \tag{S1}$$

Here, $a_i^\dagger(\hat{\mathbf{k}})$ and $a_i(\hat{\mathbf{k}})$ are the creation and annihilation operators of eigen momentum states $\hat{\mathbf{k}}$ with $\hat{\imath} = 1$ and 2, and the local coordinates are defined as

$$\hat{x}_B^{\hat{a}}(\xi'') = \int_0^{\xi''} \frac{dx^\mu}{d\xi'} e_\mu{}^{\hat{a}}\left(x^\nu(\xi')\right) d\xi'; \ \ x^\nu(\xi') = \int_0^{\xi'} \frac{dx^\nu}{d\xi} d\xi. \tag{S2}$$

Accordingly, the eigen-helicity states of photons can be described with complex scalar fields such that

$$\phi(\hat{x}) = \int \mathrm{d}^3\hat{\mathbf{k}} \, \frac{1}{\sqrt{(2\pi)^3 2k^{\hat{0}}}} \left( a(\hat{\mathbf{k}}, \sigma = -1) e^{-i\left(k_{\hat{\mu}} x^{\hat{\mu}}\right)} + a^\dagger(\hat{\mathbf{k}}, \sigma = 1) e^{i\left(k_{\hat{\mu}} x^{\hat{\mu}}\right)} \right) \tag{S3}$$

with

$$a^{\dagger}\left(\hat{\mathbf{k}},\sigma=1\right)=\frac{1}{\sqrt{2}}a_{1}^{\dagger}\left(\hat{\mathbf{k}}\right)+\frac{i}{\sqrt{2}}a_{2}^{\dagger}\left(\hat{\mathbf{k}}\right) \text{ and } a\left(\hat{\mathbf{k}},\sigma=-1\right)=\frac{1}{\sqrt{2}}a_{1}\left(\hat{\mathbf{k}}\right)+\frac{i}{\sqrt{2}}a_{2}\left(\hat{\mathbf{k}}\right).\qquad\text{(S4)}$$

Then, considering the variation of tetrad fields along geodesics of photons as well, the ansatz of complex scalar fields operator $\phi(\hat{x})$, describing the photon's helicity ($\sigma$) states, accounting for the Wigner rotation angle $\psi(\Lambda(\hat{x}),$ $\hat{\mathbf{k}}/|\hat{\mathbf{k}}|)$ are introduced such that

$$\phi\left(\hat{x}\right)=\int d^{3}\hat{\mathbf{k}}\,\frac{e^{i\left(\psi\left(\hat{x},\hat{\mathbf{k}}/|\hat{\mathbf{k}}|\right)\right)}}{\sqrt{\left(2\pi\right)^{3}2\hat{k}^{\hat{0}}}}\left(a\left(\hat{\mathbf{k}}_{\Lambda\left(\hat{x},\hat{\mathbf{k}}/|\hat{\mathbf{k}}|\right)},\sigma=-1\right)e^{-i\left(k_{\hat{\mu}}x^{\hat{a}}\right)}+a^{\dagger}\left(\hat{\mathbf{k}}_{\Lambda\left(\hat{x},\hat{\mathbf{k}}/|\hat{\mathbf{k}}|\right)},\sigma=1\right)e^{i\left(k_{\hat{\mu}}x^{\hat{a}}\right)}\right)\quad\text{(S5)}$$

Here, it is exploited that under the Lorentz transformation $\Lambda$, the creation and annihilation operators are transformed as[32]

$$U\left(\Lambda\right)a^{\dagger}\left(\hat{\mathbf{k}},\sigma\right)U^{\dagger}\left(\Lambda\right)=\sqrt{\frac{\left(\Lambda k\right)^{\hat{0}}}{k^{\hat{0}}}}e^{i\sigma\psi\left(\Lambda,\hat{\mathbf{k}}/|\hat{\mathbf{k}}|\right)}a^{\dagger}\left(\hat{\mathbf{k}}_{\Lambda},\sigma\right)$$
$$U\left(\Lambda\right)a\left(\hat{\mathbf{k}},\sigma\right)U^{\dagger}\left(\Lambda\right)=\sqrt{\frac{\left(\Lambda k\right)^{\hat{0}}}{k^{\hat{0}}}}e^{-i\sigma\psi\left(\Lambda,\hat{\mathbf{k}}/|\hat{\mathbf{k}}|\right)}a\left(\hat{\mathbf{k}}_{\Lambda},\sigma\right)$$
$$\text{(S6)}$$

As with the Berry phase, the Wigner rotation angle (WRA) can be expressed solely in terms of the local time-coordinate ($\hat{t}=x^{\hat{0}}$) along a photon's trajectory $\hat{x}(\xi)$ as

$$\psi\left(\hat{t},\hat{\mathbf{k}}/|\hat{\mathbf{k}}|\right)=\int_{0}^{\xi_{\hat{t}}}d\xi\,k^{\hat{0}}\frac{1}{k^{\hat{0}}}k^{\hat{a}}\partial_{\hat{a}}\psi\left(\Lambda\left(\hat{x}\right),\hat{\mathbf{k}}/|\hat{\mathbf{k}}|\right)=\int_{0}^{\hat{t}}d\hat{t}\,\frac{k^{\hat{a}}}{k^{\hat{0}}}\partial_{\hat{a}}\psi\left(\Lambda\left(\hat{x}\right),\hat{\mathbf{k}}/|\hat{\mathbf{k}}|\right)\qquad\text{(S7)}$$

where $\hat{t}(\xi_{\hat{t}})=\hat{t}$. Subsequently, the related field operator and Lagrangian density $\mathcal{L}$ can be formulated such that

$$\phi\left(\hat{x}\right)=\int d^{3}\hat{\mathbf{k}}\,\frac{e^{i\left(\psi\left(\hat{x},\hat{\mathbf{k}}/|\hat{\mathbf{k}}|\right)\right)}}{\sqrt{\left(2\pi\right)^{3}2\hat{k}^{\hat{0}}}}\left(a\left(\hat{\mathbf{k}}_{\Lambda\left(\hat{t},\hat{\mathbf{k}}/|\hat{\mathbf{k}}|\right)},\sigma=-1\right)e^{-i\left(k_{\hat{\mu}}x^{\hat{a}}\right)}+a^{\dagger}\left(\hat{\mathbf{k}}_{\Lambda\left(\hat{t},\hat{\mathbf{k}}/|\hat{\mathbf{k}}|\right)},\sigma=1\right)e^{i\left(k_{\hat{\mu}}x^{\hat{a}}\right)}\right)=\int d^{3}\hat{\mathbf{k}}\phi_{\hat{\mathbf{k}}}\left(\hat{x}\right)\quad\text{(S8)}$$

and

$$\mathcal{L}=\int d^{3}\hat{\mathbf{k}}\mathcal{L}_{\hat{\mathbf{k}}}=\int d^{3}\hat{\mathbf{k}}\left(D_{\hat{\mathbf{k}}}^{\ \hat{a}}\phi_{\hat{\mathbf{k}}}\left(\hat{x}\right)\right)^{\dagger}D_{\hat{\mathbf{k}}\hat{a}}\phi_{\hat{\mathbf{k}}}\left(\hat{x}\right)-j_{\hat{\mathbf{k}}}^{\ \hat{a}}A_{\hat{\mathbf{k}}\hat{a}}\qquad\text{(S9)}$$

with $D_{\hat{\mathbf{k}}\hat{a}}=\nabla_{\hat{a}}-i\partial_{\hat{a}}\psi(\hat{t},\hat{\mathbf{k}}/|\hat{\mathbf{k}}|)$. Here, $\nabla_{\hat{a}}$ is the covariant derivative defined with Christoffel symbols of a local frame, $\nabla_{\hat{c}}V^{\hat{b}}=\partial_{\hat{c}}V^{\hat{b}}+(\partial_{\hat{c}}e_{d}^{\ \mu}(\hat{x}))e_{\mu}^{\ \hat{b}}(\hat{x})V^{\hat{a}}=\partial_{\hat{c}}V^{\hat{b}}+\Gamma^{\hat{b}}_{\ \hat{c}\hat{a}}V^{\hat{a}}$. In the Lagrangian density, the term $\partial_{\hat{a}}\psi(\hat{t},\hat{\mathbf{k}}/|\hat{\mathbf{k}}|)$ can be treated as a gauge field $A_{\hat{\mathbf{k}}\hat{a}}$ interacting with a photon. With this, we derive the Euler-Lagrange equations as follows:

$$\nabla_{\hat{a}} \frac{\partial \mathcal{L}_{\mathbf{k}}}{\partial \left( \nabla_{\hat{a}} \phi_{\mathbf{k}}^{\dagger} \right)} - \frac{\partial \mathcal{L}_{\mathbf{k}}}{\partial \phi_{\mathbf{k}}^{\dagger}} = 0$$

$$\nabla_{\hat{a}} \left( \nabla^{\hat{a}} - iA_{\mathbf{k}}^{\hat{a}} \right) \phi_{\mathbf{k}} - iA_{\mathbf{k}\hat{a}} \left( \nabla^{\hat{a}} - iA_{\mathbf{k}}^{\hat{a}} \right) \phi_{\mathbf{k}} = 0 \qquad (S10)$$

$$\therefore \left( \nabla_{\hat{a}} - iA_{\mathbf{k}\hat{a}} \right) \left( \nabla^{\hat{a}} - iA_{\mathbf{k}}^{\hat{a}} \right) \phi_{\mathbf{k}} = 0$$

and

$$\nabla_{\hat{b}} \frac{\partial \mathcal{L}_{\mathbf{k}}}{\partial \left( \nabla_{\hat{b}} A_{\mathbf{k}\hat{a}} \right)} - \frac{\partial \mathcal{L}_{\mathbf{k}}}{\partial A_{\mathbf{k}\hat{a}}} = 0$$

$$j_{\mathbf{k}}^{\hat{a}} + i \left( \phi_{\mathbf{k}}^{\dagger} \left( \nabla^{\hat{a}} \phi_{\mathbf{k}} \right) - \left( \phi_{\mathbf{k}}^{\dagger} \nabla^{\hat{a}} \right) \phi_{\mathbf{k}} \right) - 2A_{\mathbf{k}}^{\hat{a}} \phi_{\mathbf{k}}^{\dagger} \phi_{\mathbf{k}} = 0$$

$$j_{\mathbf{k}}^{\hat{a}} = -i \left[ \left\{ \left( \nabla^{\hat{a}} - iA_{\mathbf{k}}^{\hat{a}} \right) \phi_{\mathbf{k}} \right\}^{\dagger} \phi_{\mathbf{k}} - \phi_{\mathbf{k}}^{\dagger} \left( \nabla^{\hat{a}} - iA_{\mathbf{k}}^{\hat{a}} \right) \phi_{\mathbf{k}} \right] \qquad (S11)$$

$$\therefore j_{\mathbf{k}}^{\hat{a}} = -i \left[ \left( D_{\mathbf{k}}^{\hat{a}} \phi_{\mathbf{k}} \right)^{\dagger} \phi_{\mathbf{k}} - \phi_{\mathbf{k}}^{\dagger} D_{\mathbf{k}}^{\hat{a}} \phi_{\mathbf{k}} \right]$$

Considering the introduced gauge field depends on the momentum, it is challenging to define a field strength tensor whose surface integration over an enclosed region becomes the line integration of Wigner rotation angle; the momentum dependence could lead to non-reciprocity, and WRAs integrated along an open path not to be cancelled out with those along the opposite direction. Therefore, we introduce the exterior-current coupled term $j\hat{\mathbf{k}}^{\hat{a}} A_{\hat{\mathbf{k}}\hat{a}}$ rather than the field strength term $F_{\hat{\mathbf{k}}\hat{a}\hat{b}} F_{\hat{\mathbf{k}}}^{\hat{a}\hat{b}}$ in the Lagrangian, treating the gauge field as an exterior background field as the case of Coriolis force and quantum Hall effect.

**Derivation of breakdown of local time-reversal symmetry using Euler-Lagrange equations**

According to the equivalence principle, in each local frame, local time reversal operator should act as time reversal operator does in quantum field theory, which transforms timelike tetrad $e^{\hat{t}}$ into $-e^{\hat{t}}$. Considering the transformation of creation and annihilation operators under the local-time reversal operator $\mathcal{T}$, $\mathcal{T} a_{\hat{t}}^{\dagger}(\hat{\mathbf{k}}) \mathcal{T}^{-1} = e^{i\zeta} a_{\hat{t}}^{\dagger}(-\hat{\mathbf{k}})$ and $\mathcal{T} a_{\hat{t}}^{\dagger}(\hat{\mathbf{k}}) \mathcal{T}^{-1} = e^{-i\zeta} a_{\hat{t}}^{\dagger}(-\hat{\mathbf{k}})$ with $\hat{t} = 1$ and $2$[32], and taking into account anti-unitarity of the operator $\mathcal{T}$ as well, the creation and annihilation operators for eigen-helicity states are transformed as follows:

$$\mathcal{T} a^{\dagger} \left( \hat{\mathbf{k}}, \sigma = 1 \right) \mathcal{T}^{-1} = \mathcal{T} \frac{\left( a_{1}^{\dagger} \left( \hat{\mathbf{k}} \right) + i a_{2}^{\dagger} \left( \hat{\mathbf{k}} \right) \right)}{\sqrt{2}} \mathcal{T}^{-1} = e^{i\zeta} \left( \frac{a_{1}^{\dagger} \left( -\hat{\mathbf{k}} \right)}{\sqrt{2}} - i \frac{a_{2}^{\dagger} \left( -\hat{\mathbf{k}} \right)}{\sqrt{2}} \right) = e^{i\zeta} a^{\dagger} \left( -\hat{\mathbf{k}}, \sigma = 1 \right),$$

$$\mathcal{T} a \left( \hat{\mathbf{k}}, \sigma = -1 \right) \mathcal{T}^{-1} = \mathcal{T} \frac{\left( a_{1} \left( \hat{\mathbf{k}} \right) + i a_{2} \left( \hat{\mathbf{k}} \right) \right)}{\sqrt{2}} \mathcal{T}^{-1} = e^{-i\zeta} \left( \frac{a_{1} \left( \hat{\mathbf{k}} \right)}{\sqrt{2}} - i \frac{a_{2} \left( -\hat{\mathbf{k}} \right)}{\sqrt{2}} \right) = e^{-i\zeta} a \left( -\hat{\mathbf{k}}, \sigma = -1 \right),$$

$(S12)$

Then, the time reversed field operator has the form

$$\mathcal{T}\phi\left(\hat{t},\hat{\mathbf{x}}\right)\mathcal{T}^{-1}=\int \mathrm{d}^3\hat{\mathbf{k}}\,\mathcal{T}\phi_{\hat{\mathbf{k}}}\left(\hat{t},\hat{\mathbf{x}}\right)\mathcal{T}^{-1}$$

$$=\int \mathrm{d}^3\hat{\mathbf{k}}\,\mathcal{T}\frac{e^{-i\left(\psi\left(\hat{t},\hat{\mathbf{k}}/|\hat{\mathbf{k}}|\right)\right)}}{\sqrt{\left(2\pi\right)^3 2\hat{k}^0}}\left(a\left(\hat{\mathbf{k}}_{\Lambda\left(\hat{t},\hat{\mathbf{k}}/|\hat{\mathbf{k}}|\right)},\sigma=-1\right)e^{-i\left(-k_{\hat{j}}\hat{t}+\hat{\mathbf{k}}_{\Lambda\left(\hat{t},\hat{\mathbf{k}}/|\hat{\mathbf{k}}|\right)}\cdot\hat{\mathbf{x}}\right)}+a^\dagger\left(\hat{\mathbf{k}}_{\Lambda\left(\hat{t},\hat{\mathbf{k}}/|\hat{\mathbf{k}}|\right)},\sigma=1\right)e^{-i\left(-k_{\hat{j}}\hat{t}+\hat{\mathbf{k}}_{\Lambda\left(\hat{t},\hat{\mathbf{k}}/|\hat{\mathbf{k}}|\right)}\cdot\hat{\mathbf{x}}\right)}\right)\mathcal{T}^{-1}$$

$$=\int \mathrm{d}^3\hat{\mathbf{k}}\,\frac{e^{-i\left(\psi\left(\hat{t},\hat{\mathbf{k}}/|\hat{\mathbf{k}}|\right)\right)}}{\sqrt{\left(2\pi\right)^3 2\hat{k}^0}}\left(a\left(-\hat{\mathbf{k}}_{\Lambda\left(\hat{t},\hat{\mathbf{k}}/|\hat{\mathbf{k}}|\right)},\sigma=-1\right)e^{i\left(-k_{\hat{j}}\hat{t}+\hat{\mathbf{k}}_{\Lambda\left(\hat{t},\hat{\mathbf{k}}/|\hat{\mathbf{k}}|\right)}\cdot\hat{\mathbf{x}}-\zeta\right)}+a^\dagger\left(-\hat{\mathbf{k}}_{\Lambda\left(\hat{t},\hat{\mathbf{k}}/|\hat{\mathbf{k}}|\right)},\sigma=1\right)e^{-i\left(-k_{\hat{j}}\hat{t}+\hat{\mathbf{k}}_{\Lambda\left(\hat{t},\hat{\mathbf{k}}/|\hat{\mathbf{k}}|\right)}\cdot\hat{\mathbf{x}}-\zeta\right)}\right)\tag{S13}$$

$$=\int \mathrm{d}^3\hat{\mathbf{k}}\,\Phi_{-\hat{\mathbf{k}}}\left(\hat{t},\hat{\mathbf{x}}\right)$$

$$=\int \mathrm{d}^3\hat{\mathbf{k}}\,\frac{e^{-i\left(\psi\left(\hat{t},-\hat{\mathbf{k}}/|\hat{\mathbf{k}}|\right)\right)}}{\sqrt{\left(2\pi\right)^3 2\hat{k}^0}}\left(a\left(\hat{\mathbf{k}}_{\Lambda\left(\hat{t},\hat{\mathbf{k}}/|\hat{\mathbf{k}}|\right)},\sigma=-1\right)e^{-i\left(k_{\hat{j}}\hat{t}+\hat{\mathbf{k}}_{\Lambda\left(\hat{t},\hat{\mathbf{k}}/|\hat{\mathbf{k}}|\right)}\cdot\hat{\mathbf{x}}-\zeta\right)}+a^\dagger\left(-\hat{\mathbf{k}}_{\Lambda\left(\hat{t},\hat{\mathbf{k}}/|\hat{\mathbf{k}}|\right)},\sigma=1\right)e^{i\left(k_{\hat{j}}\hat{t}+\hat{\mathbf{k}}_{\Lambda\left(\hat{t},\hat{\mathbf{k}}/|\hat{\mathbf{k}}|\right)}\cdot\hat{\mathbf{x}}-\zeta\right)}\right)$$

$$=\int \mathrm{d}^3\hat{\mathbf{k}}\,\Phi_{\hat{\mathbf{k}}}\left(\hat{t},\hat{\mathbf{x}}\right).$$

Here, $\Phi_{\hat{\mathbf{k}}}(\hat{x})$ represents the time-reversed field operator of mode $\hat{\mathbf{k}}$. While WRA is defined in the positive local time coordinate domain, we can extend the gauge field on the entire local time coordinate to enable the time-reversed field operator to be considered the field operator $\phi(-\hat{t},\hat{\mathbf{x}})$ with the sign-flipped local time coordinate such that

$$A_{\hat{\mathbf{k}}\hat{a}}\left(\hat{t}\right)=\begin{cases}\partial_{\hat{a}}\psi\left(\hat{t},\ \hat{\mathbf{k}}/\left|\hat{\mathbf{k}}\right|\right) & \text{for } \hat{t}\geq 0 \\ -\partial_{\hat{a}}\psi\left(-\hat{t},\ -\hat{\mathbf{k}}/\left|\hat{\mathbf{k}}\right|\right) & \text{for } \hat{t}<0\end{cases}\tag{S14}$$

Then, considering the introduced gauge field $A_{\hat{\mathbf{k}}\hat{a}}(\hat{t})$ is real scalar field but not an operator or complex function, it undergoes the transformation as

$$\mathcal{T}A_{\hat{\mathbf{k}}\hat{a}}\left(\hat{t}\right)\mathcal{T}^{-1}=\partial_{\hat{a}}\psi\left(\hat{t},\ \hat{\mathbf{k}}/\left|\hat{\mathbf{k}}\right|\right)=A_{\hat{\mathbf{k}}\hat{a}}\left(\hat{t}\right)=-A_{-\hat{\mathbf{k}}\hat{a}}\left(-\hat{t}\right).\tag{S15}$$

Then, the time reversed Lagrangian $\mathcal{T}\mathcal{L}_{\hat{\mathbf{k}}}\mathcal{T}^{-1}$ of a mode $\hat{\mathbf{k}}$ field operator becomes

$$\mathcal{T}\mathcal{L}_{\hat{\mathbf{k}}}\mathcal{T}^{-1}=\mathcal{T}\left(D_{\hat{\mathbf{k}}}^{\ \hat{a}}\phi_{\hat{\mathbf{k}}}\left(\hat{x}\right)\right)^\dagger\mathcal{T}^{-1}\mathcal{T}D_{\hat{\mathbf{k}}\hat{a}}\phi_{\hat{\mathbf{k}}}\left(\hat{x}\right)\mathcal{T}^{-1}-\mathcal{T}j_{\hat{\mathbf{k}}}^{\ \hat{a}}\mathcal{T}^{-1}\mathcal{T}A_{\hat{\mathbf{k}}\hat{a}}\mathcal{T}^{-1}$$

$$=\mathcal{T}\left(D_{\hat{\mathbf{k}}}^{\ \hat{a}}\phi_{\hat{\mathbf{k}}}\left(\hat{x}\right)\right)^\dagger\mathcal{T}^{-1}\mathcal{T}D_{\hat{\mathbf{k}}\hat{a}}\mathcal{T}^{-1}\mathcal{T}\phi_{\hat{\mathbf{k}}}\left(\hat{x}\right)\mathcal{T}^{-1}-\mathcal{T}j_{\hat{\mathbf{k}}}^{\ \hat{a}}\mathcal{T}^{-1}\mathcal{T}A_{\hat{\mathbf{k}}\hat{a}}\mathcal{T}^{-1}$$

$$=\left(\mathcal{T}\left(\nabla^{\hat{a}}\phi_{\hat{\mathbf{k}}}\left(\hat{x}\right)^\dagger-iA_{\hat{\mathbf{k}}}^{\ \hat{a}}\right)\mathcal{T}^{-1}\right)\left(\left(\nabla_{\hat{a}}+i\mathcal{T}A_{\hat{\mathbf{k}}\hat{a}}\mathcal{T}^{-1}\right)\mathcal{T}\phi_{\hat{\mathbf{k}}}\left(\hat{x}\right)\mathcal{T}^?\right)-\mathcal{T}j_{\hat{\mathbf{k}}}^{\ \hat{a}}\mathcal{T}^{-1}\mathcal{T}A_{\hat{\mathbf{k}}\hat{a}}\mathcal{T}^{-1}\tag{S16}$$

$$=\left(\left(\nabla^{\hat{a}}+iA_{\hat{\mathbf{k}}}^{\ \hat{a}}\right)\left(\mathcal{T}\phi_{\hat{\mathbf{k}}}\left(\hat{x}\right)\mathcal{T}^{-1}\right)\right)^\dagger\left(\left(\nabla_{\hat{a}}+iA_{\hat{\mathbf{k}}\hat{a}}\right)\left(\mathcal{T}\phi_{\hat{\mathbf{k}}}\left(\hat{x}\right)\mathcal{T}^{-1}\right)\right)-\mathcal{T}j_{\hat{\mathbf{k}}}^{\ \hat{a}}\mathcal{T}^{-1}\mathcal{T}A_{\hat{\mathbf{k}}\hat{a}}\mathcal{T}^{-1}.$$

Here, we use that

$$\left(\mathcal{T}\phi\left(\hat{t},\hat{\mathbf{x}}\right)\mathcal{T}^{-1}\right)^{\dagger}=\int\mathrm{d}^{3}\hat{\mathbf{k}}\left\{\frac{e^{-i\left(\psi\left(\hat{t},-\hat{\mathbf{k}}/|\hat{\mathbf{k}}|\right)\right)}}{\sqrt{\left(2\pi\right)^{3}2\hat{k}^{0}}}\left(a\left(\hat{\mathbf{k}}_{\Lambda\left(\hat{t},\hat{\mathbf{k}}/|\hat{\mathbf{k}}|\right)},\sigma=-1\right)e^{-i\left(k_{j}\hat{t}+\hat{k}_{\Lambda\left(\hat{t},\hat{\mathbf{k}}/|\hat{\mathbf{k}}|\right)}\cdot\hat{\mathbf{x}}-\zeta\right)}+a^{\dagger}\left(\hat{\mathbf{k}}_{\Lambda\left(\hat{t},\hat{\mathbf{k}}/|\hat{\mathbf{k}}|\right)},\sigma=1\right)e^{i\left(k_{j}\hat{t}+\hat{k}_{\Lambda\left(\hat{t},\hat{\mathbf{k}}/|\hat{\mathbf{k}}|\right)}\cdot\hat{\mathbf{x}}-\zeta\right)}\right)\right\}^{\dagger}$$

$$=\int\mathrm{d}^{3}\hat{\mathbf{k}}\frac{e^{i\left(\psi\left(\hat{t},-\hat{\mathbf{k}}/|\hat{\mathbf{k}}|\right)\right)}}{\sqrt{\left(2\pi\right)^{3}2\hat{k}^{0}}}\left(a^{\dagger}\left(\hat{\mathbf{k}}_{\Lambda\left(\hat{t},\hat{\mathbf{k}}/|\hat{\mathbf{k}}|\right)},\sigma=-1\right)e^{i\left(k_{j}\hat{t}+\hat{k}_{\Lambda\left(\hat{t},\hat{\mathbf{k}}/|\hat{\mathbf{k}}|\right)}\cdot\hat{\mathbf{x}}-\zeta\right)}+a\left(\hat{\mathbf{k}}_{\Lambda\left(\hat{t},\hat{\mathbf{k}}/|\hat{\mathbf{k}}|\right)},\sigma=1\right)e^{-i\left(k_{j}\hat{t}+\hat{k}_{\Lambda\left(\hat{t},\hat{\mathbf{k}}/|\hat{\mathbf{k}}|\right)}\cdot\hat{\mathbf{x}}-\zeta\right)}\right)$$

$$\text{(S17)}$$

$$=\int\mathrm{d}^{3}\hat{\mathbf{k}}\,\mathcal{T}\frac{e^{-i\left(\psi\left(\hat{t},-\hat{\mathbf{k}}/|\hat{\mathbf{k}}|\right)\right)}}{\sqrt{\left(2\pi\right)^{3}2\hat{k}^{0}}}\left(a^{\dagger}\left(-\hat{\mathbf{k}}_{\Lambda\left(\hat{t},\hat{\mathbf{k}}/|\hat{\mathbf{k}}|\right)},\sigma=-1\right)e^{-i\left(k_{j}\hat{t}+\hat{k}_{\Lambda\left(\hat{t},\hat{\mathbf{k}}/|\hat{\mathbf{k}}|\right)}\cdot\hat{\mathbf{x}}\right)}+a\left(-\hat{\mathbf{k}}_{\Lambda\left(\hat{t},\hat{\mathbf{k}}/|\hat{\mathbf{k}}|\right)},\sigma=1\right)e^{i\left(k_{j}\hat{t}+\hat{k}_{\Lambda\left(\hat{t},\hat{\mathbf{k}}/|\hat{\mathbf{k}}|\right)}\cdot\hat{\mathbf{x}}\right)}\right)\mathcal{T}^{-1}$$

$$=\int\mathrm{d}^{3}\hat{\mathbf{k}}\,\mathcal{T}\left\{\frac{e^{i\left(\psi\left(\hat{t},\hat{\mathbf{k}}/|\hat{\mathbf{k}}|\right)\right)}}{\sqrt{\left(2\pi\right)^{3}2\hat{k}^{0}}}\left(a\left(\hat{\mathbf{k}}_{\Lambda\left(\hat{t},\hat{\mathbf{k}}/|\hat{\mathbf{k}}|\right)},\sigma=-1\right)e^{i\left(-k_{j}\hat{t}+\hat{k}_{\Lambda\left(\hat{t},\hat{\mathbf{k}}/|\hat{\mathbf{k}}|\right)}\cdot\hat{\mathbf{x}}\right)}+a^{\dagger}\left(\hat{\mathbf{k}}_{\Lambda\left(\hat{t},\hat{\mathbf{k}}/|\hat{\mathbf{k}}|\right)},\sigma=1\right)e^{i\left(-k_{j}\hat{t}+\hat{k}_{\Lambda\left(\hat{t},\hat{\mathbf{k}}/|\hat{\mathbf{k}}|\right)}\cdot\hat{\mathbf{x}}\right)}\right)\right\}^{\dagger}\mathcal{T}^{-1}$$

$$=\left(\mathcal{T}\phi\left(\hat{t},\hat{\mathbf{x}}\right)^{\dagger}\mathcal{T}^{-1}\right).$$

As a result, the local time-reversed Euler-Lagrange equation of a mode $\hat{\mathbf{k}}$ field operator is given by

$$\nabla_{\hat{a}}\frac{\partial\left\{\mathcal{T}\mathcal{L}_{\hat{\mathbf{k}}}\mathcal{T}^{-1}\right\}}{\partial\left(\mathcal{T}\nabla_{\hat{a}}\phi_{\hat{\mathbf{k}}}{}^{\dagger}\mathcal{T}^{-1}\right)}-\frac{\partial\left\{\mathcal{T}\mathcal{L}_{\hat{\mathbf{k}}}\mathcal{T}^{-1}\right\}}{\partial\left(\mathcal{T}\phi_{\hat{\mathbf{k}}}{}^{\dagger}\mathcal{T}^{-1}\right)}=0,$$

$$\left\{\nabla_{\hat{a}}\left(\nabla^{\hat{a}}+i\partial^{\hat{a}}\psi_{\hat{\mathbf{k}}}\right)+i\partial_{\hat{a}}\psi_{-\hat{\mathbf{k}}}\left(\nabla^{\hat{a}}+i\partial^{\hat{a}}\psi_{\hat{\mathbf{k}}}\right)\right\}\left(\mathcal{T}\phi_{\hat{\mathbf{k}}}\left(\hat{x}\right)\mathcal{T}^{-1}\right)=0,\qquad\text{(S18)}$$

$$\therefore\left(\nabla_{\hat{a}}+i\partial^{\hat{a}}\psi_{\hat{\mathbf{k}}}\right)\left(\nabla^{\hat{a}}+i\partial^{\hat{a}}\psi_{\hat{\mathbf{k}}}\right)\left(\mathcal{T}\phi_{\hat{\mathbf{k}}}\left(\hat{x}\right)\mathcal{T}^{-1}\right)=0,$$

where $\psi\left(\hat{t},\hat{\mathbf{k}}/|\hat{\mathbf{k}}|\right)$ is denoted by $\psi_{\hat{\mathbf{k}}}$ for simplicity.

Comparing the Euler-Lagrange equation of a mode $\hat{\mathbf{k}}$ field operator, as presented in Eq. (S10), with its counterpart derived from the time-reversed Euler-Lagrange equation,

$$\int\mathrm{d}^{3}\hat{\mathbf{k}}\left(\nabla_{\hat{a}}+i\partial_{\hat{a}}\psi_{\hat{\mathbf{k}}}\right)\left(\nabla^{\hat{a}}+i\partial^{\hat{a}}\psi_{\hat{\mathbf{k}}}\right)\mathcal{T}\phi_{\hat{\mathbf{k}}}\left(\hat{x}\right)\mathcal{T}^{-1}-\int\mathrm{d}^{3}\hat{\mathbf{k}}\left(\nabla_{\hat{a}}-i\partial_{\hat{a}}\psi_{\hat{\mathbf{k}}}\right)\left(\nabla^{\hat{a}}-i\partial^{\hat{a}}\psi_{\hat{\mathbf{k}}}\right)\phi_{\hat{\mathbf{k}}}\left(\hat{x}\right)=0,\quad\text{(S19)}$$

$$\int\mathrm{d}^{3}\hat{\mathbf{k}}\left(\nabla_{\hat{a}}+i\partial_{\hat{a}}\psi_{-\hat{\mathbf{k}}}\right)\left(\nabla^{\hat{a}}+i\partial^{\hat{a}}\psi_{-\hat{\mathbf{k}}}\right)\Phi_{\hat{\mathbf{k}}}\left(\hat{x}\right)=\int\mathrm{d}^{3}\hat{\mathbf{k}}\left(\nabla_{\hat{a}}-i\partial_{\hat{a}}\psi_{\hat{\mathbf{k}}}\right)\left(\nabla^{\hat{a}}-i\partial^{\hat{a}}\psi_{\hat{\mathbf{k}}}\right)\phi_{\hat{\mathbf{k}}}\left(\hat{x}\right),\quad\text{(S20)}$$

it is found that, unless $\partial^{\hat{a}}(\psi_{-\hat{\mathbf{k}}}+\psi_{-\hat{\mathbf{k}}})$ is zero, the field operator of mode $\hat{\mathbf{k}}$, $\phi_{\hat{\mathbf{k}}}(\hat{x})$, and the time-reversed, $\Phi_{\hat{\mathbf{k}}}(\hat{x})$, follow different governing equations. In other words, the non-reciprocity in Wigner rotation angles ($\int\mathrm{d}\psi_{-\hat{\mathbf{k}}}\mathrm{d}\xi\neq\int\mathrm{d}\psi_{\hat{\mathbf{k}}}\mathrm{d}\xi$) can be seen as breaking time reversal symmetry for the fictitious observer moving along a photon trajectory.

In this framework, each observer's frame is described as local flat spacetime and the effects of its variations from curvature of spacetime are considered as gauge fields. Lars Onsager and Hendrik Casimir demonstrated that a linear, time-invariant medium slightly perturbed from thermodynamic equilibrium, with microscopic governing equations obeying time-reversal symmetry, must be a reciprocal medium[56-58]. This theorem implies that macroscopic time-reversal symmetry breakdown does not always induce non-reciprocity such as viscous fluids where the physics is reciprocal and not time reversal symmetric, but the microscopic time-reversal symmetry breakdown is required for the linear and time-invariant system to be non-reciprocal. Given the non-reciprocity of WRA, it is consistent with Onsager's theorem to time-reversal symmetry is broken down in local frames. The breakdown of local time reversal symmetry is demonstrated above by showing the transformed photon state under the time-reversed operator in Hilbert space defined at each event does not follow Lagrange equations with the untransformed gauge field. This is consistent with the case of quantum Hall effect where time reversal symmetry breaks down by treating a gauge field as an external field.

**The effects of spin angular momentum of M87\* on WRAs**

As consistent with the report of gravity Probe B, the spinning angular momentum of the Earth does not affect WRA by to any measurable order, but that of M87* leads to a WRA differences on the order of $10^{-2}$ degrees of differences as shown in Fig.S2.

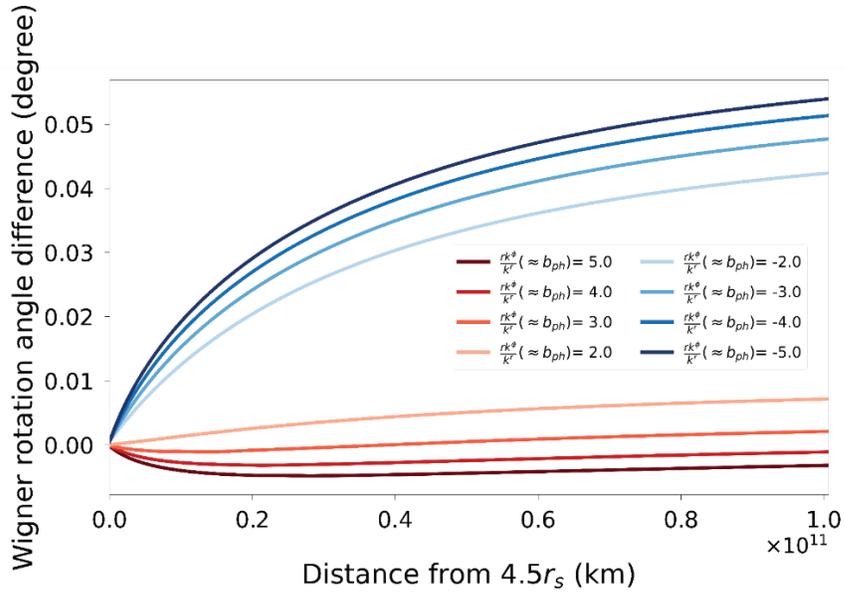

**Fig. S2** Wigner rotation angle differences between the observers orbiting around the equatorial plane around BH with the mass of M87* and zero and $0.45 r_s$ of spin parameter $a$. The large angular momentum of M87* leads to the WRA differences of the order of $10^{-2}$ degrees of for various impact factors of photons' trajectories.

**The WRAs of photon trajectories with positive and negative impact paramteres for observers in polar orbits**

In Figure S3, we present the Wigner rotation angles and $1/(1+n^3)$ for observers in polar orbits around Earth (Figs.S3A-C) and the supermassive black hole M87* (Figs.S3D-F). Notably, we observe a distinctive trend in these values for both cases. When considering negative impact parameters, $1/(1+n^3)$ exhibits values that are ten times larger compared to those associated with positive impact parameters. This significant difference in $1/(1+n^3)$ results in a consistent pattern in the Wigner rotation angles (WRAs).

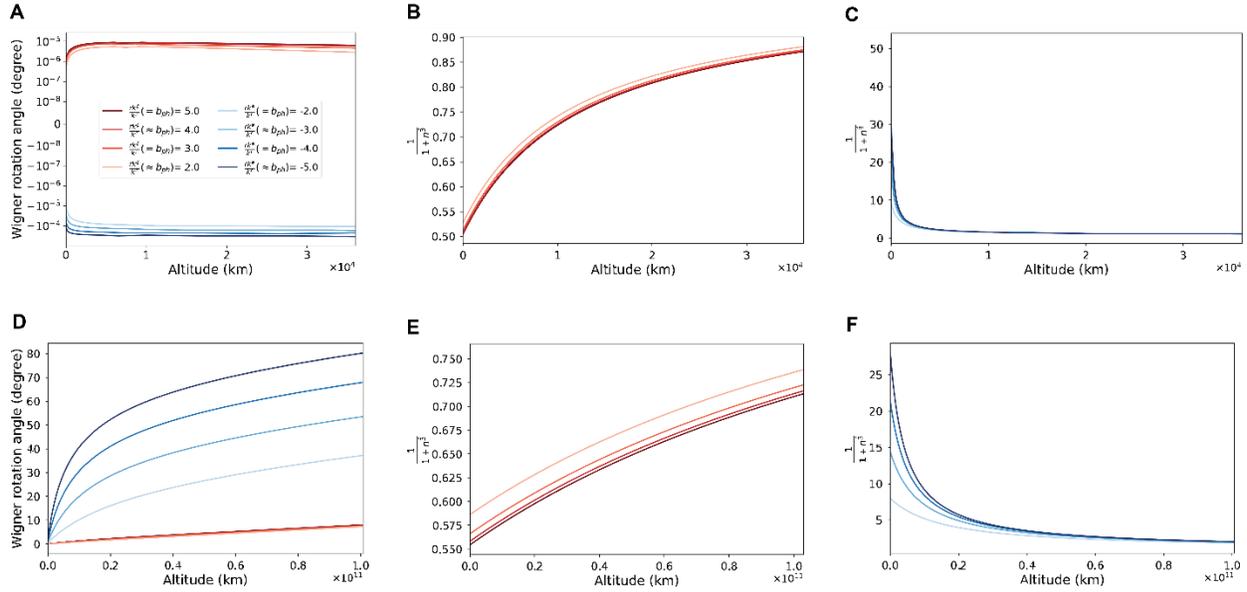

**Fig. S3** Wigner rotation angles and $1/(1+n^3)$ for the observers in polar orbits around Earth (Figs.S3A-C) and M87* (Figs.S3D-F). For both cases, $1/(1+n^3)$ has ten times larger values for negative impact parameters that those for positive, resulting in a consistent pattern in WRAs.

**WRAs under local-space inversion operator**

It is worth noting that under local-space inversion operator $\mathcal{P}$ becomes

$$\left(e_\mu{}^{\hat{i}}, e_\mu{}^{\hat{i}}\right) \underset{\mathcal{P}}{\rightarrow} \left(e_\mu{}^{\hat{i}}, -e_\mu{}^{\hat{i}}\right)$$

$$\left(\left|\hat{\mathbf{k}}\right|, \hat{\mathbf{k}}\right) \underset{\mathcal{P}}{\rightarrow} \left(k^\mu e_\mu{}^{\hat{i}}, k^\mu\left(-e_\mu{}^{\hat{i}}\right)\right) = \left(\left|\hat{\mathbf{k}}\right|, -\hat{\mathbf{k}}\right),$$

$$\nabla_k \underset{\mathcal{P}}{\rightarrow} k^\mu \nabla_\mu = \nabla_k$$

(S21)

These conditions are not obtained with flipping the signs of affine parameter $\xi$ and the proper time $\tau$ and give the opposite results of sign-change of infinitesimal Lorentz transformation terms: the signs of local infinitesimal rotations, $\lambda^{\hat{i}}_{\hat{j}} = (\nabla_k e^{\hat{j}\mu}) e_{\mu\hat{j}}$, are remain but those of local infinitesimal boosts, $\lambda^{\hat{i}}_{\hat{j}} = (\nabla_k e^{\hat{i}\mu}) e_{\mu\hat{j}} = (\nabla_k \, \mathrm{d}x^\mu/\mathrm{d}\tau) e_{\mu\hat{j}}$ , are changed. The corresponding local-space inversed infinitesimal Wigner rotation angle has the opposite sign but the same values with the local-time reversal one. Considering the sign of helicity $\sigma$ is preserved under the former only, the geometric phase $\sigma\psi_{\hat{\mathbf{k}}}$ of a photon is $\mathcal{PF}$ symmetric, but the photon trajectory defined with momentum $k$ in global coordinate is not.

**Numerical Calculation Method used for Figure 4 in the main text.**

$$\Delta\psi = \int_{\text{Path } \overline{AB}} d\psi_{\text{positive impact parameters}} - \int_{\text{Path } \overline{AC}} d\psi_{\text{negative impact parameters}}$$

$$+ \int_{\text{Path } \overline{BD}} d\psi_{\text{negative impact parameters}} - \int_{\text{Path } \overline{CD}} d\psi_{\text{positive impact parameters}}$$

$$= \int_{\text{Path } \overline{AB}} \left(d\psi_{\text{positive impact parameters}} - d\psi_{\text{negative impact parameters}}\right)$$

$$+ \int_{\text{Path } \overline{BD}} \left(d\psi_{\text{negative impact parameters}} - d\psi_{\text{positive impact parameters}}\right)$$

(S22)

Here, we utilize the understanding that the WRA near Earth depends directly on the radius '$r$', rather than on other coordinates. Based on this, the integration for the path from Alice to Bob is the same as that for Alice to Charlie. Similarly, the integration for Bob to David matches that for Charlie to David. Then, utilizing the relation $d(\Delta\psi) \equiv d\psi_{\text{positive impact parameters}} - d\psi_{\text{negative impact parameters}}$, Eq. (S22) can be reformulated as follow:

$$\Delta \psi = \int\limits_{\text{Path } \overline{AB}} d\left(\Delta \psi\right) - \int\limits_{\text{Path } \overline{BD}} d\left(\Delta \psi\right)$$

$$= \int_{r_{\text{Alice}}}^{r < \overline{AB} = \overline{A_{\text{imag}}B}} \frac{d\left(\Delta \psi\right)}{dr} dr + \int_{r \geq \overline{AB} = \overline{A_{\text{imag}}B}}^{r = \overline{OD}} \frac{d\left(\Delta \psi\right)}{dr} dr \qquad . \tag{S23}$$

Here, $r_{\text{Alice}}$ represents the distance between the origin O and Alice. The distance from Alice to David ($\overline{AB}$) needs to be determined for the calculation of the relative WRA difference WRA ($\Delta \psi$). To obtain the distance, first, we obtain the angle $\beta$ between Alice and Bob by solving the following equations for obtaining,

$$a^2 + b^2 - 2ab\cos\left(2\beta\right) = c^2 \text{ and } a^2 + c^2 - 2ac\cos\left(\pi - \alpha\right) = b^2, \tag{S24}$$

where $a$, $b$, and $\alpha$ are given. Then, by substituting the $\beta$ into the relation $h(\cot\alpha + \tan\left(\pi/2 - \alpha + 2\beta\right)) = \overline{AD}$ and then $h$ into $\overline{AB} = \overline{A_{\text{imag}}B} = h\csc\alpha$, we obtain the distance between Alice and Bob for the integration given in Eq. (S23). Here $h$ is half the distance between Bob and Charlie as depicted in Fig.S4.

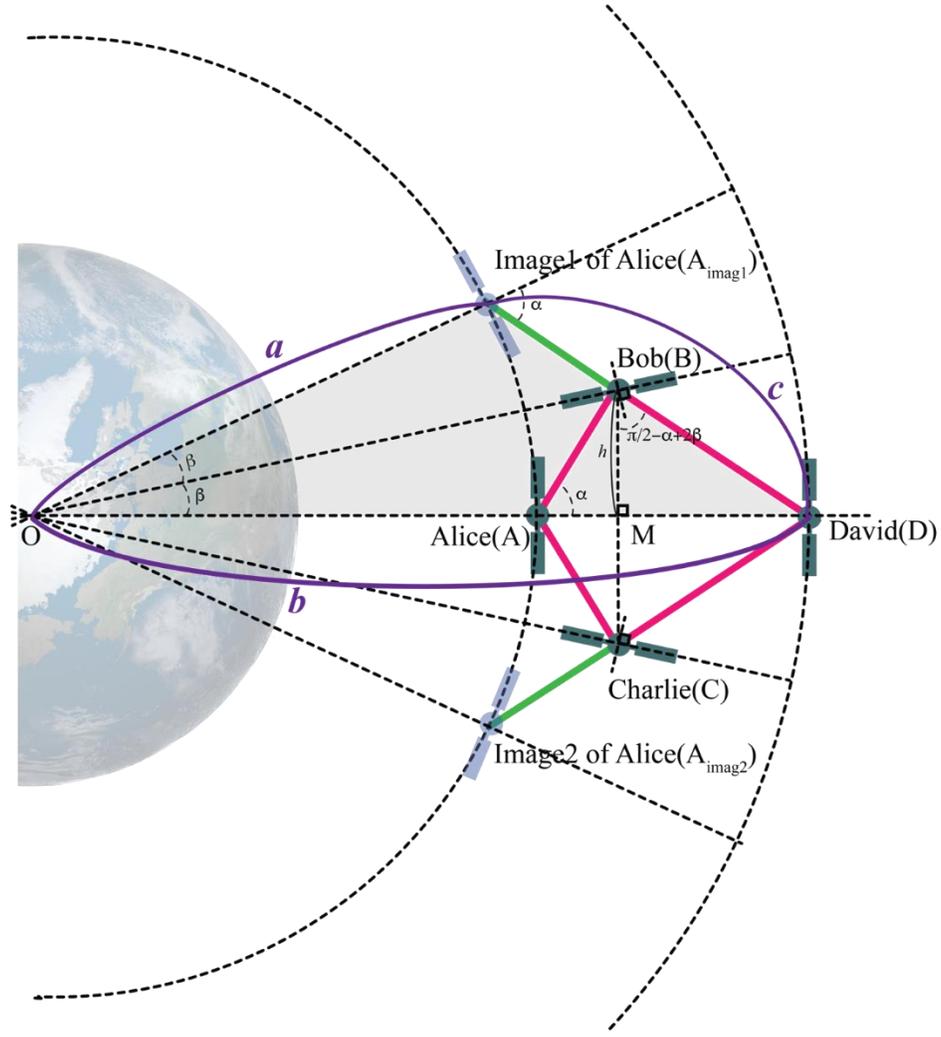

**Fig. S4** Schematic representation of an astronomical interferometer showcasing various elements, including the lengths between the origin and observers ($\overline{OA_{imag1}} = a$, $\overline{OD} = b$, $\overline{A_{imag1}D} = c$, and $\overline{BC}/2 = \overline{MB} = h$), along with the angles of outgoing light waves ($\angle MAB = \alpha$) and between two observers Alice and Bob ($\angle AOB = \beta$). The distance between Alice and Bob ($\overline{AB} = \overline{A_{imag}B} = h\csc\alpha$) is specified by the relation, $h(\cot\alpha + \tan(\pi/2 - \alpha + 2\beta)) = \overline{AD}$, $a^2 + b^2 - 2ab\cos 2\beta = c^2$, and $a^2 + c^2 - 2ac\cos(\pi - \alpha) = b^2$, proved that $\tan(\alpha)$ and $\overline{AD}$ are given.